\documentclass[sigconf]{acmart}

\AtBeginDocument{%
  }

\renewcommand\footnotetextcopyrightpermission[1]{}

\usepackage{setspace}
\usepackage{bm}
\usepackage{xcolor}
\usepackage{booktabs}
\usepackage{microtype}
\usepackage{amsfonts}
\usepackage{xspace}
\usepackage{wasysym}
\usepackage{enumitem,etoolbox}
\usepackage{array}
\usepackage{microtype}
\usepackage{multirow}
\usepackage{framed}
\usepackage{caption}
\usepackage{threeparttable}
\usepackage{appendix}
\usepackage[labelformat=simple]{subcaption}

\usepackage[ruled,vlined]{algorithm2e}

\AtBeginEnvironment{verbatim}{\setlist[trivlist]{nosep}}
\usepackage{flushend}
\usepackage{balance}
\usepackage{graphicx}

\usepackage{amsthm}
\usepackage{amsmath}
\usepackage{amssymb}

\usepackage{bbding}
\usepackage{array}
\usepackage{multirow}
\usepackage{listings}
\usepackage{color}
\usepackage{tcolorbox}
\usepackage{xcolor,colortbl}
\usepackage{enumitem}
\usepackage{adjustbox}
\usepackage[normalem]{ulem}
\usepackage{pifont}
\usepackage{tabularx}
\usepackage{subcaption}
\usepackage[capitalize]{cleveref}

\usepackage{tikz}
\usetikzlibrary{positioning}
\tikzstyle{period} = [draw=white, fill=gray!30, thick,
    rectangle, rounded corners, minimum size=9ex]
\usepackage{subcaption}
\usepackage{soul}

\usepackage{fontawesome5}

\usepackage{xcolor}
\usepackage{minted}
\usepackage{mdframed}

\definecolor{vp_bg_code}{HTML}{FAFAFA}
\definecolor{vp_bg_add}{HTML}{E6FFEC}
\definecolor{vp_bg_del}{HTML}{FFEBEB}
\definecolor{vp_bg_ref}{HTML}{E8F4FF}
\definecolor{vp_bg_doc}{HTML}{FFF8C5}
\definecolor{vp_code_black}{HTML}{2B2B2B}
\definecolor{vp_code_comment}{HTML}{6F7175}

\definecolor{acablue}{HTML}{1F77B4}   
\definecolor{acaorange}{HTML}{FF7F0E} 
\definecolor{acagreen}{HTML}{2CA02C}  
\definecolor{acared}{HTML}{D62728}    
\definecolor{acapurple}{HTML}{9467BD} 
\definecolor{acagray}{HTML}{7F7F7F}   
\definecolor{accyan}{HTML}{17BECF}    

\newcommand{\vpcodefont}{\ttfamily\fontsize{5.5pt}{6.5pt}\selectfont}

\lstdefinestyle{vp_style}{
    basicstyle=\vpcodefont\color{vp_code_black},
    keywordstyle=\bfseries\color{blue!60!black},
    commentstyle=\itshape\color{vp_code_comment},
    stringstyle=\color{red!60!black},
    numbers=left,
    numberstyle=\tiny\color{gray!50},
    stepnumber=1,
    numbersep=3pt,
    frame=none,
    escapechar=|,
    aboveskip=0pt, 
    belowskip=0pt,
    xleftmargin=1.0em, 
    breaklines=true,   
    breakatwhitespace=false,
    backgroundcolor={} 
}

\newtcolorbox{patternbox}[1]{
    colback=white,
    colframe=gray!60,
    coltitle=black,
    fonttitle=\bfseries\sffamily\fontsize{7pt}{8pt}\selectfont, 
    title={#1},
    boxrule=0.5pt,
    width=0.24\textwidth, 
    height=5.5cm,         
    arc=2pt,
    left=1pt, right=1pt, top=2pt, bottom=2pt
}

\definecolor{codegreen}{rgb}{0,0.6,0}
\definecolor{codegray}{rgb}{0.5,0.5,0.5}
\definecolor{codepurple}{rgb}{0.58,0,0.82}
\definecolor{backcolour}{HTML}{FAFAFA}
\definecolor{LightGray}{gray}{0.9}
\definecolor{LightGray2}{gray}{0.97}
\lstdefinestyle{mystyle}{
    backgroundcolor=\color{backcolour},   
    basicstyle=\vpcodefont\color{vp_code_black},
    keywordstyle=\bfseries\color{blue!60!black},
    commentstyle=\itshape\color{vp_code_comment},
    stringstyle=\color{red!60!black},
    breaklines=true,                 
    captionpos=b,                    
    keepspaces=true,                 
    showstringspaces=false,
    showtabs=false,                 
    frame = trBL,
    emph={square,root},emphstyle=\underbar,
    emph={[2]character,class,Range,Items,Item,Char,alternation,concatenation,greedy,quantifier,lazy,word,boundary,start,of,line,end,anchor,capturing,group,positive,lookahead,lookaround,negative,lookbehind,named,Character,Class,Alternation,Concatenation,Greedy,Quantifier,Lazy,Word,Boundary,Non,Start,Of,Line,End,Anchor,Capturing,Group,Positive,Lookahead,Lookaround,Negative,Lookbehind,Named,Set,Name,Integer,Star,Plus,Kleene,Optional,Repetition,Backreference},
    emphstyle={[2]\color{blue}},
    morecomment=[s][\color{codegreen}]{<}{>},
    tabsize=2
}

\usetikzlibrary{shapes,positioning}
\makeatletter
\let\old@lstKV@SwitchCases\lstKV@SwitchCases
\def\lstKV@SwitchCases#1#2#3{}
\makeatother
\usepackage{lstlinebgrd}
\makeatletter
\let\lstKV@SwitchCases\old@lstKV@SwitchCases

\lst@Key{numbers}{none}{%
    \def\lst@PlaceNumber{\lst@linebgrd}%
    \lstKV@SwitchCases{#1}%
    {none:\\%
        left:\def\lst@PlaceNumber{\llap{\normalfont
                \lst@numberstyle{\thelstnumber}\kern\lst@numbersep}\lst@linebgrd}\\%
        right:\def\lst@PlaceNumber{\rlap{\normalfont
                \kern\linewidth \kern\lst@numbersep
                \lst@numberstyle{\thelstnumber}}\lst@linebgrd}%
    }{\PackageError{Listings}{Numbers #1 unknown}\@ehc}}
\makeatother

\definecolor{rowcolor}{HTML}{ECEFF4}

\setlist[itemize]{leftmargin=*}

\definecolor{dkgreen}{rgb}{0,0.6,0}
\definecolor{gray}{rgb}{0.5,0.5,0.5}
\definecolor{mauve}{rgb}{0.58,0,0.82}

\newcommand{\eg}{\hbox{\emph{e.g.}}\xspace}
\newcommand{\ie}{\hbox{\emph{i.e.}}\xspace}

\newcommand{\bench}{\textsc{LiveFMBench}\xspace}

\setlist[itemize]{leftmargin=*}
\setlist[enumerate]{leftmargin=*}
\newlist{steps}{enumerate}{1}
\setlist[steps, 1]{label = \textbf{RQ\arabic*.}}

\begin{document}

\title{LiveFMBench: Unveiling the Power and Limits of Agentic Workflows in Specification Generation
}

\author{Dong Xu}
\email{xudong2022@iscas.ac.cn}
\affiliation{
  \institution{Institute of Software, Chinese Academy of Sciences}
  \city{Beijing}
  \country{China}
}
\author{Jialun Cao}
\email{jialuncao@ust.hk}
\affiliation{
  \institution{The Hong Kong University of Science and Technology}
  \city{Hong Kong}
  \country{China}
}
\author{Guozhao Mo}
\email{moguozhao2024@iscas.ac.cn}
\affiliation{
  \institution{Institute of Software, Chinese Academy of Sciences}
  \city{Beijing}
  \country{China}
}
\author{Junjie Hu}
\email{24181214428@stu.xidian.edu.cn}
\affiliation{
  \institution{Guangzhou Institute of Technology, Xidian University}
  \city{Guangzhou}
  \country{China}
}
\author{Cheng Wen}
\email{wencheng@xidian.edu.cn}
\affiliation{
  \institution{Guangzhou Institute of Technology, Xidian University}
  \city{Guangzhou}
  \country{China}
}
\author{Hongyu Lin}
\email{hongyu@iscas.ac.cn}
\affiliation{
  \institution{Institute of Software, Chinese Academy of Sciences}
  \city{Beijing}
  \country{China}
}
\author{Xianpei Han}
\email{xianpei@iscas.ac.cn}
\affiliation{
  \institution{Institute of Software, Chinese Academy of Sciences}
  \city{Beijing}
  \country{China}
}
\author{Shengchao Qin}
\email{shengchao.qin@gmail.com}
\affiliation{
  \institution{Guangzhou Institute of Technology, Xidian University}
  \city{Guangzhou}
  \country{China}
}
\author{Cong Tian}
\email{ctian@mail.xidian.edu.cn}
\affiliation{
  \institution{Guangzhou Institute of Technology, Xidian University}
  \city{Guangzhou}
  \country{China}
}
\author{Shing-Chi Cheung}
\email{scc@cse.ust.hk}
\affiliation{
  \institution{Institute of Software, Chinese Academy of Sciences}
  \city{Beijing}
  \country{China}
}
\author{Le Sun}
\email{sunle@iscas.ac.cn}
\affiliation{
  \institution{Institute of Software, Chinese Academy of Sciences}
  \city{Beijing}
  \country{China}
}
\author{Yaojie Lu}
\email{luyaojie@iscas.ac.cn}
\affiliation{
  \institution{Institute of Software, Chinese Academy of Sciences}
  \city{Beijing}
  \country{China}
}
\renewcommand{\shortauthors}{Xu et al.}

\begin{abstract}

Formal specification is essential for rigorous program verification, yet writing correct specifications remains costly and difficult to automate. Although large language models (LLMs) and agents have shown promising progress, their true capabilities and failure modes remain unclear. We present the first systematic and contamination-aware study of LLM- and agent-based formal specification generation for C programs. We introduce \bench, a continuously evolving benchmark of 630 ACSL (ANSI/ISO C Specification Language)-annotated C programs, including 360 newly collected cases designed to mitigate data leakage. Using this benchmark, we evaluate direct prompting with different sampling sizes, reasoning-enabled (thinking mode) inference, the agentic pipeline, and perform a fine-grained failure analysis.
Experimental results reveal that naive evaluation substantially overestimates performance because models under direct prompting may exhibit unfaithful behaviors, such as deceiving automated provers or ignoring code-context constraints; after excluding such cases, the true specification generation accuracy drops by approximately 20\%. 
We further find that both increased sampling and thinking mode significantly improve success rates, with smaller models benefiting more from thinking mode. 
Agentic pipelines are particularly effective under low sampling budgets and on harder datasets. Failure analysis further shows that incorrect loop invariants are the dominant error type, while agentic pipelines notably reduce assertion errors.
These results expose fundamental limitations in current LLM-based approaches and suggest they remain far from replacing human-authored formal specifications. We release \bench at \url{https://huggingface.co/datasets/fm-universe/Live-FM-Bench} and all evaluation artifacts to support future research.

\end{abstract}

\maketitle

\section{Introduction}\label{sec:intro}

Formal program verification provides a rigorous foundation for establishing critical program properties, such as functional correctness, safety, and liveness. 
Despite its strong theoretical guarantees and immense industrial demand~\cite{ebalard2019journey,efremov2018deductive,dordowsky2015experimental,blanchard2015case,kosmatov2014case}, practical verification remains challenging due to the difficulty of constructing correct specifications~\cite{hahnle2019deductive,code2inv18,wen2024enchantingprogramspecificationsynthesis}. 

Recently, large language models (LLMs) have demonstrated impressive capabilities in code understanding~\cite{nam2024using,wang2025empirical,richards2024you,yan-etal-2024-codescope} and generation~\cite{gu2025retrieve,llm-code-gen-survey-in-low-resource,humaneval,mpbb2021}, motivating a growing body of work that applies them to formal specification and program verification tasks~\cite{SpecGen25,first2023baldur,endres2023formalizing,pei2023can,kogler2024reliable,wang2023lego-prover}. 
Prior work has explored LLM-assisted formal verification across a diverse set of formalisms, including interactive theorem provers such as Coq~\cite{coqgym,kasibatla2026cobblestone} and Lean~\cite{lean4,yang2023leandojo,ying2024lean,song2024towards,hsiang2025leandojo,leanexpert24,tao2023formalizinglean4}, program verification languages such as Dafny~\cite{mugnier2024laurelgeneratingdafnyassertions,LLM4Dafny,leino2010dafny}, temporal logic–based specification frameworks such as TLA+~\cite{zhou2025towards,zhou2025retrieval}, as well as contract-based specification languages like ACSL (ANSI/ISO C Specification Language)~\cite{wen2024enchantingprogramspecificationsynthesis}.
These works suggested promising results and thus raised optimism about the practical applicability of AI-assisted formal verification
~\cite{kasibatla2026cobblestone,proofcoop26,wen2024enchantingprogramspecificationsynthesis,SpecGen25,ma2025bridging}.

However, despite the encouraging progress, existing results should be interpreted with caution. 
First, \textit{Data leakage threatens the validity}. Most benchmarks for the evaluation were collected from GitHub or contests, which are likely to overlap with the pre-training data of LLMs, making it difficult to disentangle genuine reasoning from memorization. 
Second, most existing works concentrate on applying AI techniques (such as prompt engineering, in-context learning~\cite{wen2024enchantingprogramspecificationsynthesis}, retrieval-augmented generation~\cite{gu2025retrieve}, and multi-round iterative refinement~\cite{wen2024enchantingprogramspecificationsynthesis,SpecGen25}) to the verification of programs written in different programming languages (\eg, C, Rust, and Java), \textbf{offering limited insight into how and why LLM succeed or fail}. As a result, the underlying limitations of LLM reasoning, semantic understanding, and specification modeling remain largely unexplored. Third, although recent work increasingly adopts agent-based designs~\cite{wen2024enchantingprogramspecificationsynthesis} or explicit reasoning prompts~\cite{5team2025glm45agenticreasoningcoding,deepseekai2025deepseekr1incentivizingreasoningcapability}, the benefits of these techniques are often demonstrated in isolated settings, \textbf{without systematic analysis} across different reasoning modes, program complexities, or failure scenarios.
In particular, failures in specification generation are often subtle: an incomplete precondition, an unsound postcondition, or an overlooked frame condition may invalidate subsequent verification, even if the generated specification appears superficially plausible.

To support the study, we introduce \bench, a continuously evolving benchmark for formal specification generation and verification. \bench consists of 630 verifiable C programs annotated with ACSL specifications~\cite{acsl_reference}, designed to enable both rigorous evaluation and long-term reuse while being aware of mitigating the data contamination issue. Using this benchmark, we conduct a comprehensive empirical study to progressively investigate three core questions:

\textbf{First, Direct Prompting of LLMs.} 
While prior work has primarily focused on designing pipelines~\cite{wen2024enchantingprogramspecificationsynthesis,SpecGen25} or fine-tuning LLMs~\cite{fmbench} to improve task-specific performance, it remains unclear \textbf{\textit{whether the intrinsic capabilities of LLMs have been fully exploited}}. 
To investigate this question, we quantify the effect of increased sampling, assess the faithfulness of the existing evaluation process, and measure the benefits of reasoning-enabled thinking mode. Essentially, this question explores whether the inherent capabilities of current LLMs have been fully exploited, beyond carefully engineered pipelines or fine-tuning.

\textbf{Second, Benefits from the agentic pipeline.} 
Assuming the LLMs' intrinsic capabilities are near their limits, we further investigate \textit{\textbf{whether agentic orchestration provides additional benefits}}. Recent approaches~\cite{wen2024enchantingprogramspecificationsynthesis,tale-1001} increasingly adopt multi-step agents for formal verification tasks, but it is still unclear which agent components contribute most to performance gains, and whether reasoning-enabled thinking mode is beneficial or detrimental within these agent components. 

\textbf{Third, Causes of failure.} 
Even under best-effort conditions, LLM-based and Agentic-based (abbreviated as \textit{AI} for brevity) still fail to produce correct formal specifications, which motivates us to dig into {\textit{why LLM fails in specification generation}}.
We perform a systematic analysis of failure cases, categorizing error types, measuring their prevalence, and investigating correlations with program properties and model behavior. Furthermore, we examine model familiarity with programs and program-plus-specification pairs to provide empirical indicators for whether fine-tuning may be necessary for improved generalization. 

\textit{\textbf{Findings}} -- Our study yields several important insights into the reliability and performance of LLM-based formal specification generation. 
\textbf{First, at the reliability level}, models under direct prompting exhibit non-negligible unfaithful behaviors, such as deceiving automated provers or ignoring constraints from the code context. These behaviors can substantially inflate apparent performance: after filtering out such deceptive cases, the measured accuracy drops by approximately 20\%. 
\textbf{Second, at the model performance level}, we observe five consistent performance trends: 
1) Increasing the sample size yields substantial gains, with pass@5 \textbf{doubling} pass@1 and pass@32 \textbf{tripling} it on average; 
2) Thinking mode consistently improves performance, with relative gains ranging from 19.40\% to 2465.52\%, indicating that explicit reasoning can greatly enhance specification generation performance;
3) Smaller models benefit more from thinking mode, as illustrated by Qwen3-32B improving from 6.33 to 27.44 on pass@5.
4) The agentic pipeline is particularly effective under low sampling budgets and on harder datasets, although its advantage diminishes as sampling increases.
\textbf{Third, the failure analysis} shows that incorrect loop invariants are the most common failure type in both reasoning modes, while the agentic pipeline notably reduces assertion errors.

Overall, our study shows that although LLMs can generate well-formed and plausible formal specifications, their understanding of deep semantic dependencies, implicit constraints, and global program behavior remains limited. As program complexity grows, these weaknesses become more pronounced, leading to brittle and unreliable specifications. 
In summary, this paper makes the following contributions:

\begin{itemize}
    \item \textbf{\textit{Benchmark}}. We introduce \textbf{\bench}, the first continuously updatable and contamination-aware benchmark for formal specification generation, consisting of 630 verifiable C programs annotated with ACSL specifications. The benchmark includes a temporally stratified split with newly collected programs to enable contamination-resistant evaluation of LLM generalization.
    
    \item \textbf{\textit{Systematic Study}}. We conduct the first comprehensive empirical study of LLM- and agent-based formal specification generation under direct prompt, reasoning-enabled (thinking mode), and agentic pipeline settings. Our evaluation framework corrects common evaluation biases and provides a principled measurement of true specification generation capability.
    
    \item \textbf{\textit{Insight}}. We present a fine-grained failure taxonomy and quantitative root-cause analysis, revealing why current models fail in formal specification generation. Our analysis identifies dominant error sources such as missing specifications, incorrect pre/postconditions, flawed loop invariants, and failures in capturing implicit assumptions and non-local program semantics.
    
    \item \textbf{\textit{Reproducibility}}. We release \bench, all evaluation scripts, annotations, and failure labels to support reproducible, extensible, and contamination-aware research on AI-assisted formal verification.
\end{itemize}

\section{Study Design}
This section describes the research questions (Section~\ref{sec:rqs}), formal specification selection, and task selection.

\begin{figure*}[t!]
    \centering
    \includegraphics[width=0.95\linewidth]{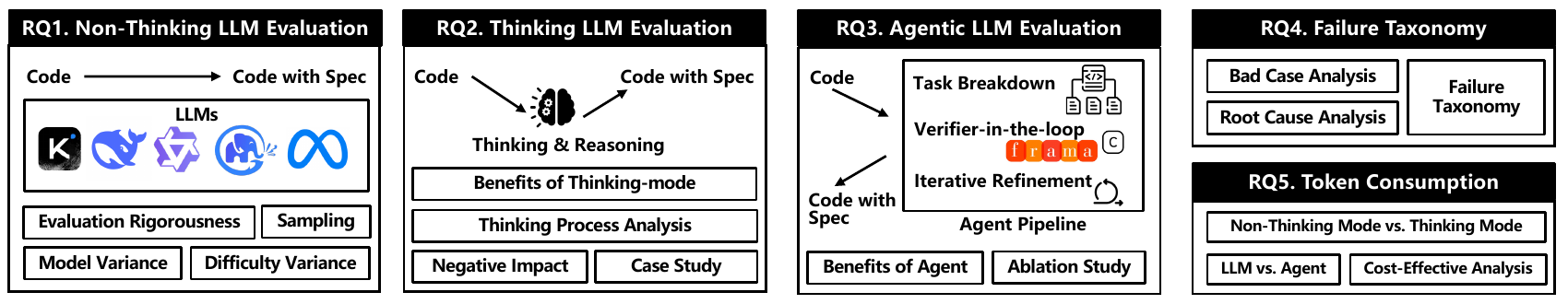}
    \caption{Study Design for RQ1-RQ5}
    \label{fig:design}
\end{figure*}

\subsection{Research Questions}\label{sec:rqs}
To explore the three core questions raised in Section~\ref{sec:intro}, we design five research questions (RQs):

\begin{itemize}
    \item \textbf{\textit{RQ1. How well do direct prompting LLMs perform in formal specification generation?}} We first explore LLMs' non-thinking mode in a direct prompting manner, \ie, given a program under verification and the property to be verified, checking whether the output program and specifications can satisfy the property (\ie, passing the verifier) or not. We also increased the number of samples (\ie, Pass@1, Pass@5, Pass@32) to explore how far LLMs can improve. Finally, we closely examine the outputs and assess whether there are gaps in the evaluation. 
    
    \item \textbf{\textit{RQ2. How far can thinking mode improve formal specification generation?}} 
Recent advances in LLMs have introduced explicit \emph{thinking modes}, where models generate intermediate reasoning steps before producing final answers. Such reasoning-enabled models have been widely observed to outperform direct-generation baselines across a range of complex tasks, including mathematical problem solving, code generation and program repair. Motivated by these gains, we investigate whether and to what extent thinking mode can improve formal specification generation. Specifically, we compare thinking and non-thinking settings in terms of verification success rate (specification correctness) and further analyze how the benefits of thinking mode vary across programs of different difficulty levels. 

\item  \textbf{\textit{RQ3. How much can an agent pipeline improve specification generation?}} Beyond directly prompting, agentic pipelines that orchestrate multiple reasoning steps, such as iterative refinement and verifier-guided feedback, were also proposed. Motivated by these advances, we investigate whether and to what extent the agent can improve the specification generation. We evaluate the performance gains brought by static analysis, verifier-in-the-loop feedback, and self-refinement, measuring improvements in verification success rate. Furthermore, we analyze how agent-based approaches scale across programs of different complexity levels, and examine whether they can systematically reduce common failure modes. Finally, we conduct a qualitative analysis to understand when the agent provides substantial benefits and when it introduces additional overhead or new sources of error.

\item \textbf{\textit{RQ4. Why do LLMs fail in formal specification generation?}} 
Then, we conduct a systematic root-cause analysis of failed specification generation cases. We categorize errors into fine-grained failure types, such as missing specifications (e.g., absent preconditions, postconditions, or loop invariants), incorrect functional constraints, unsound loop invariants, and verifier-specific misuse. We then quantify the prevalence of each failure category, providing a statistical breakdown of the dominant error sources. Furthermore, we analyze how failure patterns differ across model configurations, thinking modes, agent pipelines, and program difficulty levels. Through this analysis, we aim to uncover the fundamental bottlenecks in current AI-driven formal specification generation and highlight directions for future improvement.

\item \textbf{\textit{RQ5. How much tokens do LLMs cost in formal specification generation?}} 
Finally, we analyze the token consumption of LLMs and its relationship with accuracy in different inference modes and approaches. Test-time scaling \cite{snell2024scaling} refers to the paradigm of allocating additional computation during the inference phase (i.e. spend more tokens) to boost a model's performance. This is typically achieved by providing richer context information (e.g. via agent pipelines), enabling multiple generation attempts (such as optimizing for $pass@k$ with a large $k$), and leveraging explicit reasoning chains to systematically increase the overall pass rate. We first investigated whether these three test-time scaling methods could improve model performance in writing ACSL specifications, and subsequently analyzed which method is more cost-effective under a limited token budget.
\end{itemize}

\subsection{Formal Specification Selection}
Formal specification languages vary in their intended use and strengths: Proof assistants such as Isabelle and Coq~\cite{coqgym} are widely used in recent LLM studies, but they primarily target the verification of mathematical theorems rather than direct program behavior. Formal languages like Dafny~\cite{LLM4Dafny,mugnier2024laurelgeneratingdafnyassertions} and Lean~\cite{lean4,yang2023leandojo,ying2024lean,song2024towards,hsiang2025leandojo,leanexpert24,tao2023formalizinglean4} provide program verification support, but often impose restrictions or require complex annotations that make them less practical for large-scale C program verification. Model-checking languages such as TLA+~\cite{cousineau2012tlaproofs,zhou2025towards,zhou2025retrieval} and others, focus on abstract system behavior and are not designed to capture the detailed semantics of concrete program execution.

In contrast, ACSL~\cite{acsl_reference} is specifically tailored for specifying and verifying C programs. It allows precise expression of function contracts, memory safety properties, and data invariants, and integrates well with existing verification tools such as Frama-C. Importantly, recent studies indicate that large language models achieve better performance on ACSL than on other specification languages such as Coq or Lean4, likely due to ACSL’s closer alignment with typical C code patterns.

By choosing ACSL, we aim to reduce the noise that may arise from a model’s unfamiliarity with the specification language, ensuring that our evaluation reflects the model’s ability to generate correct program specifications rather than its general reasoning about formal languages. Therefore, ACSL is the natural choice for our study of automated specification generation for C programs.

\subsection{Task Selection}
The automated generation of ACSL from C programs and localized assertions is a critical research frontier driven by two foundational challenges in software engineering.

First, it significantly lowers the high barrier to entry for formal verification. While developers routinely write simple executable assertions for debugging, constructing complete formal contracts—such as precise pre/post-conditions and mathematically sound loop invariants—demands specialized expertise in first-order logic. Automating this process enables developers to leverage rigorous deductive verification tools using only familiar C code artifacts.

Second, it addresses the theoretical difficulty of deriving global program specifications from localized intent anchors. A simple assertion merely checks an isolated intermediate state. To mathematically prove its validity across all execution paths, a system must infer the function's complete logical closure. This requires complex logical abduction to synthesize overarching guarantees and accurate loop invariants, making the Code-to-ACSL task a rigorous benchmark for deep semantic program comprehension.

\section{Benchmark Construction}\label{sec:bench}

\begin{figure*}[hbt!]
    \centering
    \includegraphics[width=0.95\linewidth]{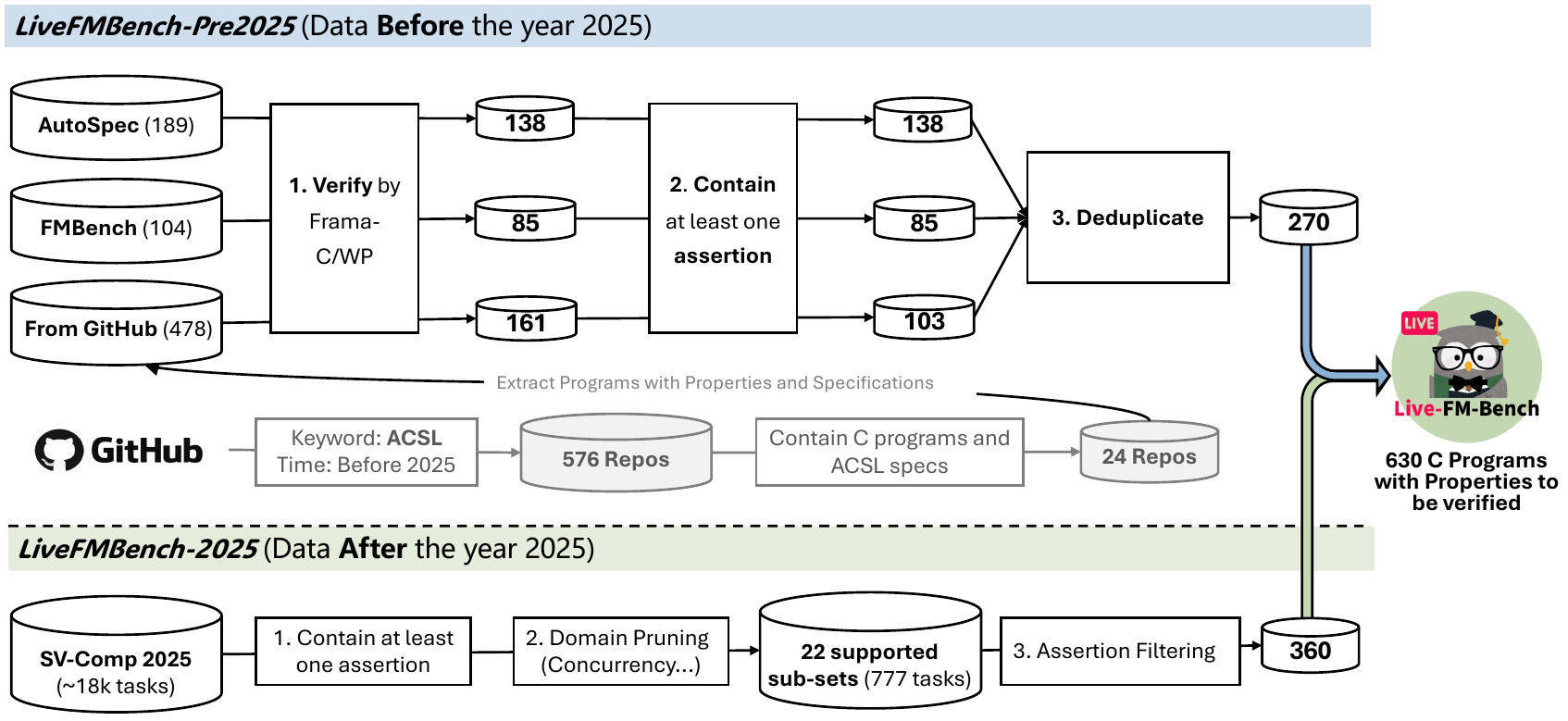}
    \caption{Data Collection and Preparation for \bench}
    \label{fig:data-prepare}
\end{figure*}

The collection and preparation pipeline of \bench~ is illustrated in Figure~\ref{fig:data-prepare}. 
It consists of two subsets based on temporal boundaries: (1) \textbf{\bench-pre2025} (Section~\ref{sec:data-pre25}), which may have been included in model training corpora, obtained from legacy programs in existing benchmarks or GitHub repositories \textit{before the year 2025}; and (2) \textbf{\bench-2025} (Section~\ref{sec:data25}), which was collected \textit{in the year 2025} from the SV-COMP (\ie, International Competition on Software Verification)~\cite{sv-comp}. We then explain the process in detail. 

\subsection{\bench-pre2025: Historical Data}\label{sec:data-pre25}
As shown in the upper part of Figure~\ref{fig:data-prepare}, the \bench-re2025 is curated from \textbf{three sources}: (1) AutoSpec~\cite{wen2024enchantingprogramspecificationsynthesis}(189 programs), (2) FMBench~\cite{fmbench} (104 programs), and (3) GitHub repositories~(478 programs) that contain C code with ACSL specifications. For GitHub data collection, we searched 576 repositories with the keyword ``\texttt{ACSL}'' and filtered them to retain 24 repositories containing verifiable C programs with properties and specifications. Specifically, the data undergoes a three-step filtering process:

\begin{enumerate}
    \item \textbf{Verifiable by Frama-C/WP}: All programs are verified using Frama-C/WP~\cite{Cuoq-frama-c} to ensure syntactic correctness and compatibility with the verifier. After verification, 138, 85, and 161 programs remain from AutoSpec, FMBench, and GitHub, respectively;
    \item \textbf{Assertion Presence:} Programs are then filtered to retain only those containing \textit{at least one property} under verification (annotated by \texttt{assertion}), resulting in 138, 85, and 103 programs from each source;
    \item \textbf{Deduplication.} Duplicate programs across sources are removed, yielding a total of \textbf{270 unique C programs} in the \bench-pre2025. The deduplication was performed by computing cosine similarity between two programs (excluding code comments and specifications) using the widely used UniXcoder embedding model~\cite{guo2022unixcoder}. Programs with a cosine similarity greater than 90\% are treated as duplicates and subsequently removed from the dataset.
\end{enumerate}

\subsection{\bench-2025}\label{sec:data25}
To mitigate potential training data contamination, we collected 2025 data from the SV-COMP 2025~\cite{sv-comp}, which includes approximately 18,000 verification tasks. As shown in the lower part of Figure~\ref{fig:data-prepare}, the data construction involves three steps:

\begin{enumerate}
    \item \textbf{Assertion Presence:} 
    We retain programs containing \textit{at least one property} under verification annotated by \texttt{assertion};
    \item \textbf{Domain Pruning:} Verification tasks involving unsupported domains (e.g., concurrency) are removed, resulting in 777 programs across 22 supported subdomains that span classical verification themes such as array reasoning, heap manipulation, and nested loops.;
    \item \textbf{Assertion Filtering:} We further filter to select programs with suitable assertions for ACSL-based verification, producing 360 unique programs in \bench-2025.
    
\end{enumerate}

\subsection{An Example Program and Its Specifications in \bench}
Listing~\ref{lst:example} shows a case from \bench, which includes (1) C programs (white-background code); (2) property to be verified (\ie, \texttt{//@ assert a[i].n == 0;}) in line 24, annotated in green; (3) ACSL specifications, \eg, \texttt{loop invariant} annotated in yellow. 

To verify this program, one should annotate the C program with formal logic (pre-conditions, post-conditions, invariants) written in ACSL specification language, then use deductive verification tools like Frama-C~\cite{Cuoq-frama-c} with its WP (\ie, Weakest Precondition) plug-in~\cite{frama-c-wp}. The WP plugin generates mathematical proof obligations that are automatically or interactively verified by solvers like Z3 and Coq. 

\definecolor{color-green}{HTML}{E6F1EA}
\definecolor{color-red}{HTML}{faccd0}
\definecolor{color-yellow}{HTML}{FCF3D5}

\colorlet{FancyVerbHighlightColor}{color-yellow}
\begin{listing}[]
\begin{minted}{cpp}
#include <stdlib.h>
#include <assert.h>
#define SIZE 100000
typedef struct{
  int *n;
}S;
void init(S a[], int size){
  int i;
  /*@
    loop invariant \forall integer j; 0 <= j < i ==> a[j].n != \null || a[j].n == \null;
    loop invariant 0 <= i;
  */
  for(i = 0; i < size; i++)  {
    a[i].n = (int *) malloc(sizeof(int));
  }
}
int main(){
  S a[SIZE];
  int i;
  int flag;
  // ... [Codes and specs are omitted for brevity] ...
  for(i = 0; i < SIZE; i++){
    if (flag == 0){
    |\xglobal\colorlet{FancyVerbHighlightColor}{color-green}|//@ assert a[i].n == 0; // |\textbf{\textcolor{black}{Property to be verified}}|
    }
  }
  return 0;
}  
\end{minted}
\caption{\textbf{An example of a C program with the \colorbox{color-green}{property to be verified} and \colorbox{color-yellow}{ACSL Specification} in \bench~}}
\label{lst:example}
\end{listing}

\begin{figure}[tbp] 
     \centering
     \begin{subfigure}[b]{0.7\linewidth}
         \centering
         \includegraphics[width=\textwidth]{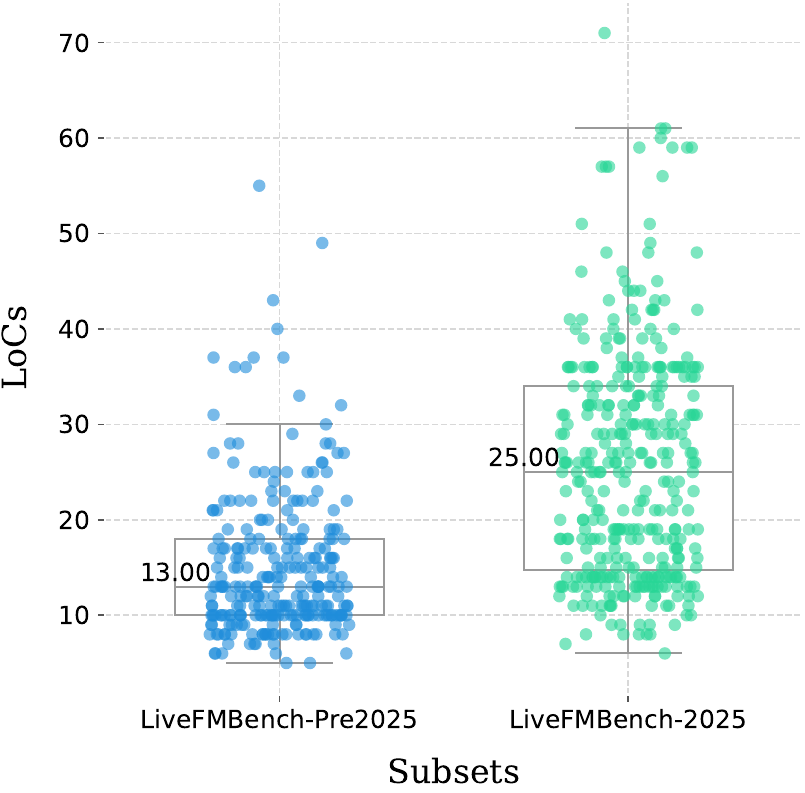}
         \caption{\centering Distribution of LoCs 
         }
         \label{fig:scatter}
     \end{subfigure}
     \hfill
  
     \begin{subfigure}[b]{0.7\linewidth}
         \centering
         \includegraphics[width=\textwidth]{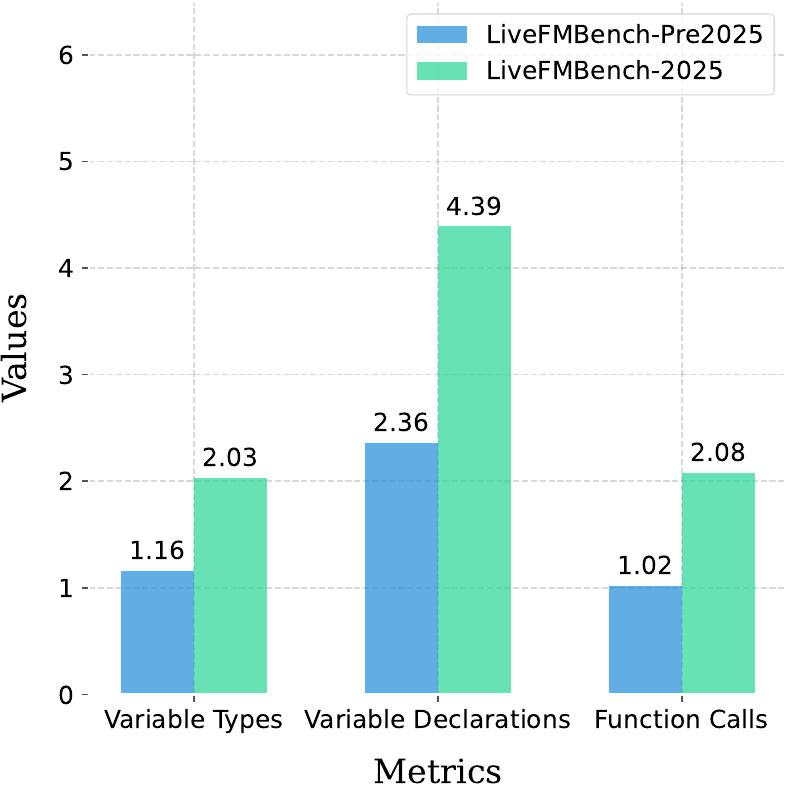}
         \caption{\centering Average of Other Metrics}
         \label{fig:bar}
     \end{subfigure}
     
     \caption{Statistics of two subsets (\colorbox[HTML]{75ADE0}{\bench-pre2025} and \colorbox[HTML]{88DFB8}{\bench-2025}) in \bench}
     \label{fig:overall_difficulty_comparison}
\end{figure}

\subsection{Dataset Statistics}

\subsubsection{Difficulty}
\label{subsubsec:difficulty}
We employ four static metrics to quantify task difficulty: Lines of Code (LOC), the number of distinct variable types, the number of variable declarations, and the total number of function calls.

As illustrated in Figure~\ref{fig:scatter}, the 2025 subset exhibits a broader distribution range compared to the 2024 subset. Across four of the five difficulty indicators—with the exception of the number of function definitions—the mean values for the 2025 subset are consistently higher. This indicates \textbf{a higher overall difficulty level}, primarily attributed to the more complex C programs introduced in the SV-COMP 2025 competition, which demand superior formal verification capabilities.

A detailed breakdown of the data distribution is provided in Figure \ref{fig:bar}. In this scatter plot, colors distinguish the two subsets, while the marker size represents the LOC. The x and y axes denote the variable complexity index (sum of types and declarations) and the function complexity index (sum of definitions and calls), respectively. Since these metrics are discrete integers, many data points overlap on a grid; therefore, the color intensity (opacity) represents the density of points at each coordinate. The 2025 subset shows a higher concentration of data points in the top-right quadrant and a wider overall dispersion, further corroborating the conclusions drawn from the violin plots.

\section{Experiment}

\subsection{Experiment Setup}

\textbf{Evaluated Models}. 
We evaluate \textbf{15 open-source LLMs}, categorized by their inference modes. The 9 non-thinking models (direct generation) include the DeepSeek-V3 (R1) family \cite{deepseekai2025deepseekv3technicalreport}, Qwen3 family \cite{yang2025qwen3technicalreport, qwen2.5-1m, hui2024qwen2}, GLM-4 family \cite{5team2025glm45agenticreasoningcoding}, and Llama-3.3-70B-Instruct \cite{llama3modelcard}. The 6 thinking models (integrating explicit internal reasoning) comprise the DeepSeek-R1 family \cite{deepseekai2025deepseekr1incentivizingreasoningcapability}, specific Qwen3 and GLM-4 models, and GPT-oss-120B. Several models, such as Qwen3-32B, Deepseek-V3.1-Terminus/V3.2, and GLM-4.6-355B-A22B, support both modes and are evaluated under both categories.

\textbf{Pipelines}.
We employed two approaches for generating specifications for C programs: (1) \textbf{Direct prompting}: LLMs directly generate specifications based on instructions and code under verification. (2) \textbf{Agentic pipeline}: Following AutoSpec~\citep{wen2024enchantingprogramspecificationsynthesis}, we construct an agentic workflow that parses execution-ordered insertion points, sequentially prompts the model for specification generation, and filters out invalid or redundant outputs.

\textbf{Agentic Setup}. For the agent-based specification generation, we adopted the mainstream AutoSpec~\cite{wen2024enchantingprogramspecificationsynthesis} framework for evaluation, establishing agent baselines by selecting four backbone models from the list in direct prompting, including Deepseek-R1-0528\cite{deepseekai2025deepseekr1incentivizingreasoningcapability}, Qwen3-32B(non-thinking mode)\cite{yang2025qwen3technicalreport}, Qwen3-32B(thinking mode)\cite{yang2025qwen3technicalreport}, and Llama3.3-70B-Instruct\cite{llama3modelcard}.

\textbf{Evaluation Metrics}. We adopt pass@$k$ as our evaluation metric, measuring the probability that at least one valid specification is generated within $k$ attempts for each C program. Aligning with established protocols in formal theorem proving, we set our maximum $k$ to 32. To minimize redundant testing and variance for lower $k$ values, we estimate these metrics directly from pass@32 using the unbiased estimator proposed by Chen et al.\cite{chen2021evaluatinglargelanguagemodels}, defined as: $\text{pass@$k$} = \mathop{\mathbb{E}}_{\text{Problem}} \left[1 - \frac{{\binom{n-c}{k}}} {\binom{n}{k}}\right]$
where $n$ is the number of total samples (\ie $n=32$, in our experiments), and $c$ is the number of samples that pass the verification. 

\textbf{Verifier Selection}.  We adopted Frama-C (v27.1)~\cite{Cuoq-frama-c, wen2024enchantingprogramspecificationsynthesis}, supported by Alt-Ergo and Z3 as the underlying provers.

\textbf{Prompts} In direct prompting, we concatenated the instructions and code directly as input to intuitively evaluate the models' raw capabilities following existing work~\cite{fmbench}. For the agentic pipeline, we reused the official open-source code and retained its prompt template from existing work~\cite{wen2024enchantingprogramspecificationsynthesis}. 

\textbf{Hyperparameters}. Following DeepSeek-R1 technical report \cite{deepseekai2025deepseekr1incentivizingreasoningcapability}, we set the temperature to 0.7, the upper bound of the recommended 0.5–0.7 range. This configuration facilitates the stochastic diversity required by the pass@32 metric to effectively capture the performance ceiling of model reasoning. Additionally, top-p was fixed at the suggested 0.95. 

\subsection{RQ1. Direct Prompting and Improvement from Sampling} \label{sec:rq1}
In this section, we analyze the performance of direct-prompting LLMs on \bench from four perspectives:  faithfulness, subset-level, metric-level, and model-level differences.

\subsubsection{Faithfulness}

We observed that in the direct-prompting pipeline, the complete program integrated with ACSL specifications generated by models frequently exhibits semantic non-equivalence with the original input program. The model may modify portions of the original program code or the assertions to be proved.  Examples are demonstrated in Listing \ref{lst:failure-faith-code} and Listing \ref{lst:failure-faith-spec} of Section \ref{subsubsec:case-study}. We compare the Abstract Syntax Trees (AST) of the generated code and the original code, and parse all ACSL assertion expressions from the original code. Only if the two ASTs are equivalent and every assertion expression in the original code appears in the generated code, is the generated code considered faithful.
\begin{table*}[tbh!]
\centering
\caption{ \centering RQ1. How well do direct prompting LLMs perform in formal specification
generation? \\ 
\small {\faRobot~ denotes the non-thinking model (inf. mode), \faLightbulb~ denotes the thinking model (inf. mode), the same below} \\
\small{Subtitles "Before" and "After" correspond to pass@k before adding faith check and after faithfulness check, respectively}
}
\label{tab:rq1}
\renewcommand\arraystretch{1.2}
    \resizebox{1.0\linewidth}{!}{
\begin{tabular}{l|rrr|rrr|rrr|rrr|rrr|rrr}
\toprule
 & \multicolumn{9}{c|}{\textbf{\bench-Pre2025}} & \multicolumn{9}{c}{\textbf{\bench-2025}} \\ \cline{2-19} 
 & \multicolumn{3}{c|}{{{\ul \textbf{Pass@1}}}} & \multicolumn{3}{c|}{{{\ul \textbf{Pass@5}}}} & \multicolumn{3}{c|}{{{\ul \textbf{Pass@32}}}} & \multicolumn{3}{c|}{{{\ul \textbf{Pass@1}}}} & \multicolumn{3}{c|}{{{\ul \textbf{Pass@5}}}} & \multicolumn{3}{c}{{{\ul \textbf{Pass@32}}}} \\
\multirow{-3}{*}{LLM} & \multicolumn{1}{c}{Before} & \multicolumn{1}{c}{After} & \multicolumn{1}{c|}{\textbf{$\Downarrow$ (\%)}} & \multicolumn{1}{c}{Before} & \multicolumn{1}{c}{After} & \multicolumn{1}{c|}{\textbf{$\Downarrow$ (\%)}} & \multicolumn{1}{c}{Before} & \multicolumn{1}{c}{After} & \multicolumn{1}{c|}{\textbf{$\Downarrow$ (\%)}} & \multicolumn{1}{c}{Before} & \multicolumn{1}{c}{After} & \multicolumn{1}{c|}{\textbf{$\Downarrow$ (\%)}} & \multicolumn{1}{c}{Before} & \multicolumn{1}{c}{After} & \multicolumn{1}{c|}{\textbf{$\Downarrow$ (\%)}} & \multicolumn{1}{c}{Before} & \multicolumn{1}{c}{After} & \multicolumn{1}{c}{\textbf{$\Downarrow$ (\%)}} \\
\midrule
\faRobot~Kimi-K2-0905 & \cellcolor[HTML]{DCF3EF}30.44 & \cellcolor[HTML]{E0F4F1}27.49 & \multicolumn{1}{c|}{\cellcolor[HTML]{F9F9F9}9.69} & \cellcolor[HTML]{C8EBE6}47.96 & \cellcolor[HTML]{CAECE7}46.02 & \multicolumn{1}{c|}{\cellcolor[HTML]{FDFDFD}4.05} & \cellcolor[HTML]{B5E4DD}64.44 & \cellcolor[HTML]{B8E6DF}61.48 & \cellcolor[HTML]{FDFDFD}4.59 & \cellcolor[HTML]{F4FBFA}9.64 & \cellcolor[HTML]{F8FDFC}6.32 & \multicolumn{1}{c|}{\cellcolor[HTML]{E9E9E9}34.44} & \cellcolor[HTML]{E6F6F4}21.95 & \cellcolor[HTML]{F0FAF9}13.04 & \multicolumn{1}{c|}{\cellcolor[HTML]{E6E6E6}40.59} & \cellcolor[HTML]{D1EFEA}40.28 & \cellcolor[HTML]{E4F6F3}23.33 & \cellcolor[HTML]{E5E5E5}42.08 \\
\faRobot~DeepSeek-V3-0324 & \cellcolor[HTML]{E3F5F3}24.29 & \cellcolor[HTML]{E9F7F5}19.47 & \multicolumn{1}{c|}{\cellcolor[HTML]{F3F3F3}19.84} & \cellcolor[HTML]{D0EEEA}40.68 & \cellcolor[HTML]{D8F1ED}34.11 & \multicolumn{1}{c|}{\cellcolor[HTML]{F5F5F5}16.15} & \cellcolor[HTML]{C0E8E2}54.81 & \cellcolor[HTML]{C9ECE7}46.67 & \cellcolor[HTML]{F6F6F6}14.85 & \cellcolor[HTML]{FDFFFE}2.06 & \cellcolor[HTML]{FEFFFF}1.28 & \multicolumn{1}{c|}{\cellcolor[HTML]{E7E7E7}37.86} & \cellcolor[HTML]{FAFEFD}4.71 & \cellcolor[HTML]{FCFEFE}2.80 & \multicolumn{1}{c|}{\cellcolor[HTML]{E6E6E6}40.55} & \cellcolor[HTML]{F4FBFA}9.72 & \cellcolor[HTML]{FAFDFD}5.00 & \cellcolor[HTML]{E0E0E0}48.56 \\
\faRobot~DeepSeek-V3.1-Terminus & \cellcolor[HTML]{DBF2EF}31.33 & \cellcolor[HTML]{E3F5F2}24.64 & \multicolumn{1}{c|}{\cellcolor[HTML]{F2F2F2}21.35} & \cellcolor[HTML]{C9ECE6}47.37 & \cellcolor[HTML]{D3EFEB}38.30 & \multicolumn{1}{c|}{\cellcolor[HTML]{F3F3F3}19.15} & \cellcolor[HTML]{B8E6DF}61.85 & \cellcolor[HTML]{C4EAE4}51.11 & \cellcolor[HTML]{F4F4F4}17.36 & \cellcolor[HTML]{FBFEFD}4.06 & \cellcolor[HTML]{FCFEFE}2.87 & \multicolumn{1}{c|}{\cellcolor[HTML]{EDEDED}29.31} & \cellcolor[HTML]{F5FCFB}9.11 & \cellcolor[HTML]{F8FDFC}6.32 & \multicolumn{1}{c|}{\cellcolor[HTML]{ECECEC}30.63} & \cellcolor[HTML]{ECF8F6}17.22 & \cellcolor[HTML]{F2FBF9}11.67 & \cellcolor[HTML]{EBEBEB}32.23 \\
\faRobot~DeepSeek-V3.2 & \cellcolor[HTML]{D5F0EC}37.00 & \cellcolor[HTML]{E5F6F3}23.24 & \multicolumn{1}{c|}{\cellcolor[HTML]{E8E8E8}37.19} & \cellcolor[HTML]{B8E6DF}61.35 & \cellcolor[HTML]{CEEDE9}43.08 & \multicolumn{1}{c|}{\cellcolor[HTML]{ECECEC}29.78} & \cellcolor[HTML]{A9E0D8}74.44 & \cellcolor[HTML]{B9E6DF}60.74 & \cellcolor[HTML]{F4F4F4}18.40 & \cellcolor[HTML]{FBFEFD}3.79 & \cellcolor[HTML]{FDFFFE}2.13 & \multicolumn{1}{c|}{\cellcolor[HTML]{E3E3E3}43.80} & \cellcolor[HTML]{F6FCFB}8.29 & \cellcolor[HTML]{FAFEFD}4.76 & \multicolumn{1}{c|}{\cellcolor[HTML]{E4E4E4}42.58} & \cellcolor[HTML]{EEF9F8}15.00 & \cellcolor[HTML]{F6FCFB}8.33 & \cellcolor[HTML]{E3E3E3}44.47 \\
\faRobot~Qwen3-Coder-480B-A35B-Instruct & \cellcolor[HTML]{DCF3EF}30.37 & \cellcolor[HTML]{E0F4F1}27.04 & \multicolumn{1}{c|}{\cellcolor[HTML]{F8F8F8}10.96} & \cellcolor[HTML]{CFEEE9}41.75 & \cellcolor[HTML]{D4F0EC}37.56 & \multicolumn{1}{c|}{\cellcolor[HTML]{F9F9F9}10.04} & \cellcolor[HTML]{C3EAE4}52.22 & \cellcolor[HTML]{CAECE7}45.93 & \cellcolor[HTML]{F8F8F8}12.05 & \cellcolor[HTML]{F9FDFD}5.43 & \cellcolor[HTML]{FAFEFD}4.55 & \multicolumn{1}{c|}{\cellcolor[HTML]{F5F5F5}16.21} & \cellcolor[HTML]{F3FBFA}11.14 & \cellcolor[HTML]{F4FBFA}9.56 & \multicolumn{1}{c|}{\cellcolor[HTML]{F6F6F6}14.18} & \cellcolor[HTML]{ECF9F7}16.39 & \cellcolor[HTML]{EFF9F8}14.44 & \cellcolor[HTML]{F8F8F8}11.90 \\
\faRobot~GLM-4.6-355B-A32B & \cellcolor[HTML]{E8F7F5}20.19 & \cellcolor[HTML]{ECF8F6}17.15 & \multicolumn{1}{c|}{\cellcolor[HTML]{F6F6F6}15.06} & \cellcolor[HTML]{D0EEEA}41.36 & \cellcolor[HTML]{D4F0EC}37.53 & \multicolumn{1}{c|}{\cellcolor[HTML]{FAFAFA}9.26} & \cellcolor[HTML]{B9E6DF}61.11 & \cellcolor[HTML]{BDE7E1}57.41 & \cellcolor[HTML]{FCFCFC}6.05 & \cellcolor[HTML]{F6FCFB}8.49 & \cellcolor[HTML]{F7FCFC}7.37 & \multicolumn{1}{c|}{\cellcolor[HTML]{F7F7F7}13.19} & \cellcolor[HTML]{E5F6F3}22.99 & \cellcolor[HTML]{E8F7F5}20.65 & \multicolumn{1}{c|}{\cellcolor[HTML]{F9F9F9}10.18} & \cellcolor[HTML]{D3EFEB}38.61 & \cellcolor[HTML]{D6F0ED}35.83 & \cellcolor[HTML]{FBFBFB}7.20 \\
\faRobot~Qwen3-235B-A22B-Instruct-2507 & \cellcolor[HTML]{EDF9F7}15.83 & \cellcolor[HTML]{F2FBF9}11.54 & \multicolumn{1}{c|}{\cellcolor[HTML]{EEEEEE}27.10} & \cellcolor[HTML]{DFF4F1}27.60 & \cellcolor[HTML]{E7F7F4}21.35 & \multicolumn{1}{c|}{\cellcolor[HTML]{F1F1F1}22.64} & \cellcolor[HTML]{D2EFEB}39.26 & \cellcolor[HTML]{DBF2EF}31.11 & \cellcolor[HTML]{F2F2F2}20.76 & \cellcolor[HTML]{FBFEFE}3.53 & \cellcolor[HTML]{FDFFFE}1.89 & \multicolumn{1}{c|}{\cellcolor[HTML]{E2E2E2}46.46} & \cellcolor[HTML]{F5FCFB}9.33 & \cellcolor[HTML]{F9FDFD}5.52 & \multicolumn{1}{c|}{\cellcolor[HTML]{E5E5E5}40.84} & \cellcolor[HTML]{EBF8F6}17.50 & \cellcolor[HTML]{F4FBFA}10.00 & \cellcolor[HTML]{E4E4E4}42.86 \\
\faRobot~Llama3.3-70B-Instruct & \cellcolor[HTML]{F2FBF9}11.64 & \cellcolor[HTML]{FBFEFD}4.27 & \multicolumn{1}{c|}{\cellcolor[HTML]{D7D7D7}63.32} & \cellcolor[HTML]{E4F6F3}23.57 & \cellcolor[HTML]{F3FBFA}10.83 & \multicolumn{1}{c|}{\cellcolor[HTML]{DDDDDD}54.05} & \cellcolor[HTML]{D2EFEB}38.89 & \cellcolor[HTML]{E8F7F5}20.37 & \cellcolor[HTML]{E1E1E1}47.62 & \cellcolor[HTML]{FBFEFE}3.56 & \cellcolor[HTML]{FFFFFF}0.25 & \multicolumn{1}{c|}{\cellcolor[HTML]{C4C4C4}92.98} & \cellcolor[HTML]{F3FBFA}10.48 & \cellcolor[HTML]{FFFFFF}0.82 & \multicolumn{1}{c|}{\cellcolor[HTML]{C5C5C5}92.18} & \cellcolor[HTML]{E5F6F3}23.06 & \cellcolor[HTML]{FCFEFE}2.78 & \cellcolor[HTML]{C7C7C7}87.94 \\
\faRobot~Qwen3-32B & \cellcolor[HTML]{F2FAF9}12.00 & \cellcolor[HTML]{F8FDFC}6.33 & \multicolumn{1}{c|}{\cellcolor[HTML]{E1E1E1}47.25} & \cellcolor[HTML]{E8F7F5}20.03 & \cellcolor[HTML]{F0FAF8}13.77 & \multicolumn{1}{c|}{\cellcolor[HTML]{EBEBEB}31.25} & \cellcolor[HTML]{DAF2EE}32.22 & \cellcolor[HTML]{E2F5F2}25.19 & \cellcolor[HTML]{F2F2F2}21.82 & \cellcolor[HTML]{FEFFFF}1.01 & \cellcolor[HTML]{FFFFFF}0.29 & \multicolumn{1}{c|}{\cellcolor[HTML]{D2D2D2}71.29} & \cellcolor[HTML]{FBFEFD}3.85 & \cellcolor[HTML]{FEFFFF}1.31 & \multicolumn{1}{c|}{\cellcolor[HTML]{D5D5D5}65.97} & \cellcolor[HTML]{F2FAF9}11.94 & \cellcolor[HTML]{F9FDFD}5.28 & \cellcolor[HTML]{DCDCDC}55.78 \\ \hline
Average & \cellcolor[HTML]{E4F6F3}23.68 & \cellcolor[HTML]{EBF8F6}17.91 & \multicolumn{1}{c|}{\cellcolor[HTML]{F0F0F0}24.37} & \cellcolor[HTML]{D2EFEB}39.07 & \cellcolor[HTML]{DBF2EF}31.39 & \multicolumn{1}{c|}{\cellcolor[HTML]{F3F3F3}19.65} & \cellcolor[HTML]{C2E9E3}53.25 & \cellcolor[HTML]{CCEDE8}44.45 & \cellcolor[HTML]{F5F5F5}16.53 & \cellcolor[HTML]{FAFEFD}4.62 & \cellcolor[HTML]{FCFEFE}2.99 & \multicolumn{1}{c|}{\cellcolor[HTML]{E9E9E9}35.17} & \cellcolor[HTML]{F2FBFA}11.32 & \cellcolor[HTML]{F7FCFC}7.20 & \multicolumn{1}{c|}{\cellcolor[HTML]{E8E8E8}36.40} & \cellcolor[HTML]{E7F7F4}21.08 & \cellcolor[HTML]{F0FAF9}12.96 & \cellcolor[HTML]{E7E7E7}38.51 \\ \hline
\faLightbulb~DeepSeeek-R1-0528 & \cellcolor[HTML]{CCEDE8}44.33 & \cellcolor[HTML]{D7F1ED}34.69 & \multicolumn{1}{c|}{\cellcolor[HTML]{F2F2F2}21.75} & \cellcolor[HTML]{A5DFD6}77.78 & \cellcolor[HTML]{AFE2DB}69.63 & \multicolumn{1}{c|}{\cellcolor[HTML]{F9F9F9}10.48} & \cellcolor[HTML]{94D8CE}92.96 & \cellcolor[HTML]{97DAD0}90.00 & \cellcolor[HTML]{FDFDFD}3.18 & \cellcolor[HTML]{E8F7F5}20.24 & \cellcolor[HTML]{EDF9F7}15.73 & \multicolumn{1}{c|}{\cellcolor[HTML]{F1F1F1}22.28} & \cellcolor[HTML]{CFEEE9}41.60 & \cellcolor[HTML]{DBF2EF}31.81 & \multicolumn{1}{c|}{\cellcolor[HTML]{F0F0F0}23.53} & \cellcolor[HTML]{B4E4DD}65.00 & \cellcolor[HTML]{C8ECE6}47.50 & \cellcolor[HTML]{EEEEEE}26.92 \\
\faLightbulb~DeepSeek-V3.1-Terminus & \cellcolor[HTML]{D3EFEB}38.48 & \cellcolor[HTML]{DCF3EF}30.52 & \multicolumn{1}{c|}{\cellcolor[HTML]{F2F2F2}20.69} & \cellcolor[HTML]{ADE2DA}71.42 & \cellcolor[HTML]{B7E5DE}62.71 & \multicolumn{1}{c|}{\cellcolor[HTML]{F8F8F8}12.20} & \cellcolor[HTML]{96D9CF}91.11 & \cellcolor[HTML]{99DAD1}88.52 & \cellcolor[HTML]{FEFEFE}2.84 & \cellcolor[HTML]{EDF9F7}16.08 & \cellcolor[HTML]{F0FAF8}13.64 & \multicolumn{1}{c|}{\cellcolor[HTML]{F6F6F6}15.17} & \cellcolor[HTML]{DEF3F0}28.81 & \cellcolor[HTML]{E2F5F2}25.00 & \multicolumn{1}{c|}{\cellcolor[HTML]{F7F7F7}13.22} & \cellcolor[HTML]{CDEDE9}43.33 & \cellcolor[HTML]{D2EFEB}39.17 & \cellcolor[HTML]{F9F9F9}9.60 \\
\faLightbulb~DeepSeek-V3.2 & \cellcolor[HTML]{CDEDE8}43.55 & \cellcolor[HTML]{DCF3EF}30.31 & \multicolumn{1}{c|}{\cellcolor[HTML]{ECECEC}30.40} & \cellcolor[HTML]{A9E0D8}74.73 & \cellcolor[HTML]{BAE6E0}59.93 & \multicolumn{1}{c|}{\cellcolor[HTML]{F3F3F3}19.80} & \cellcolor[HTML]{97DAD0}90.00 & \cellcolor[HTML]{A0DDD4}82.22 & \cellcolor[HTML]{FAFAFA}8.64 & \cellcolor[HTML]{F8FDFC}6.52 & \cellcolor[HTML]{FAFEFD}4.65 & \multicolumn{1}{c|}{\cellcolor[HTML]{EDEDED}28.68} & \cellcolor[HTML]{ECF8F7}16.79 & \cellcolor[HTML]{F0FAF9}13.09 & \multicolumn{1}{c|}{\cellcolor[HTML]{F1F1F1}22.04} & \cellcolor[HTML]{DCF3EF}30.83 & \cellcolor[HTML]{E2F5F2}25.83 & \cellcolor[HTML]{F5F5F5}16.22 \\
\faLightbulb~GLM-4.6-355B-A32B & \cellcolor[HTML]{DAF2EE}32.55 & \cellcolor[HTML]{E0F4F1}27.08 & \multicolumn{1}{c|}{\cellcolor[HTML]{F5F5F5}16.80} & \cellcolor[HTML]{B0E3DB}68.51 & \cellcolor[HTML]{B7E5DE}62.46 & \multicolumn{1}{c|}{\cellcolor[HTML]{FAFAFA}8.83} & \cellcolor[HTML]{95D9CF}91.85 & \cellcolor[HTML]{96D9CF}91.11 & \cellcolor[HTML]{FFFFFF}0.81 & \cellcolor[HTML]{EFFAF8}14.03 & \cellcolor[HTML]{F1FAF9}12.30 & \multicolumn{1}{c|}{\cellcolor[HTML]{F8F8F8}12.33} & \cellcolor[HTML]{DBF2EF}31.27 & \cellcolor[HTML]{E0F4F1}27.42 & \multicolumn{1}{c|}{\cellcolor[HTML]{F8F8F8}12.31} & \cellcolor[HTML]{C8EBE6}48.06 & \cellcolor[HTML]{CEEEE9}42.78 & \cellcolor[HTML]{F8F8F8}10.99 \\
\faLightbulb~Qwen3-235B-A22B-Thinking-2507 & \cellcolor[HTML]{D4F0EC}37.11 & \cellcolor[HTML]{D7F1ED}34.92 & \multicolumn{1}{c|}{\cellcolor[HTML]{FCFCFC}5.90} & \cellcolor[HTML]{B6E5DE}63.33 & \cellcolor[HTML]{B8E6DF}61.51 & \multicolumn{1}{c|}{\cellcolor[HTML]{FEFEFE}2.87} & \cellcolor[HTML]{A3DED5}79.63 & \cellcolor[HTML]{A4DED6}78.89 & \cellcolor[HTML]{FFFFFF}0.93 & \cellcolor[HTML]{F6FCFB}8.21 & \cellcolor[HTML]{F7FCFC}7.38 & \multicolumn{1}{c|}{\cellcolor[HTML]{F9F9F9}10.11} & \cellcolor[HTML]{E5F6F3}23.07 & \cellcolor[HTML]{E7F7F4}20.92 & \multicolumn{1}{c|}{\cellcolor[HTML]{FAFAFA}9.32} & \cellcolor[HTML]{CEEEE9}42.78 & \cellcolor[HTML]{D2EFEB}38.89 & \cellcolor[HTML]{FAFAFA}9.09 \\
\faLightbulb~Qwen3-32B & \cellcolor[HTML]{DAF2EE}32.37 & \cellcolor[HTML]{E0F4F1}27.44 & \multicolumn{1}{c|}{\cellcolor[HTML]{F6F6F6}15.23} & \cellcolor[HTML]{BBE7E0}58.92 & \cellcolor[HTML]{C3EAE4}52.16 & \multicolumn{1}{c|}{\cellcolor[HTML]{F8F8F8}11.47} & \cellcolor[HTML]{A3DED5}79.63 & \cellcolor[HTML]{AAE1D9}73.33 & \cellcolor[HTML]{FAFAFA}7.91 & \cellcolor[HTML]{F5FCFB}9.04 & \cellcolor[HTML]{F7FCFC}7.44 & \multicolumn{1}{c|}{\cellcolor[HTML]{F4F4F4}17.70} & \cellcolor[HTML]{E4F6F3}23.62 & \cellcolor[HTML]{E9F7F5}19.78 & \multicolumn{1}{c|}{\cellcolor[HTML]{F5F5F5}16.26} & \cellcolor[HTML]{D3EFEB}38.61 & \cellcolor[HTML]{D7F1ED}34.72 & \cellcolor[HTML]{F9F9F9}10.08 \\ \hline
Average & \cellcolor[HTML]{D3F0EB}38.07 & \cellcolor[HTML]{DCF3EF}30.83 & \multicolumn{1}{c|}{\cellcolor[HTML]{F3F3F3}19.02} & \cellcolor[HTML]{AFE2DB}69.12 & \cellcolor[HTML]{B8E6DF}61.40 & \multicolumn{1}{c|}{\cellcolor[HTML]{F8F8F8}11.16} & \cellcolor[HTML]{9ADBD1}87.53 & \cellcolor[HTML]{9EDCD3}84.01 & \cellcolor[HTML]{FDFDFD}4.02 & \cellcolor[HTML]{F1FAF9}12.35 & \cellcolor[HTML]{F4FBFA}10.19 & \multicolumn{1}{c|}{\cellcolor[HTML]{F4F4F4}17.51} & \cellcolor[HTML]{E0F4F1}27.53 & \cellcolor[HTML]{E5F6F3}23.00 & \multicolumn{1}{c|}{\cellcolor[HTML]{F5F5F5}16.43} & \cellcolor[HTML]{CCEDE8}44.77 & \cellcolor[HTML]{D3EFEB}38.15 & \cellcolor[HTML]{F6F6F6}14.79 \\ 

\bottomrule
\end{tabular}
}
\end{table*}
\begin{table}[thb!]
\centering
\caption{RQ1. Unfaithful Rate. Higher values are worse. 
}
\label{tab:code-change}
\renewcommand\arraystretch{1.0}
    \resizebox{1.0\linewidth}{!}{
\begin{tabular}{llrcc}
\toprule
\multicolumn{1}{c}{} &  &  & \multicolumn{2}{c}{\textbf{Changed Rate (\%)}} \\ \cline{4-5} 
\multicolumn{1}{c}{\multirow{-2}{*}{\textbf{LLM}}} & \multirow{-2}{*}{\textbf{Mode}} & \multirow{-2}{*}{\textbf{size}} & \multicolumn{1}{c}{\textbf{Pre2025}} & \multicolumn{1}{c}{\textbf{2025}} \\ \hline
Kimi-K2-0905 & Non-thinking & 1T & \cellcolor[HTML]{E8E8E8}13.54 & \cellcolor[HTML]{C8C8C8}43.06 \\
DeepSeek-V3-0324 & Non-thinking & 671B & \cellcolor[HTML]{D5D5D5}20.75 & \cellcolor[HTML]{E6E6E6}24.03 \\
DeepSeek-V3.1-Terminus & Non-thinking & 671B & \cellcolor[HTML]{D4D4D4}21.09 & \cellcolor[HTML]{D6D6D6}30.19 \\
DeepSeek-V3.2 & Non-thinking & 671B & \cellcolor[HTML]{C3C3C3}37.11 & \cellcolor[HTML]{C1C1C1}48.88 \\
Qwen3-Coder-480B-A35B-Instruct & Non-thinking & 480B & \cellcolor[HTML]{E0E0E0}15.17 & \cellcolor[HTML]{C8C8C8}42.50 \\
GLM-4.6-355B-A32B & Non-thinking & 355B & \cellcolor[HTML]{F0F0F0}12.00 & \cellcolor[HTML]{F6F6F6}20.09 \\
Qwen3-235B-A22B-Instruct-2507 & Non-thinking & 235B & \cellcolor[HTML]{D1D1D1}23.91 & \cellcolor[HTML]{BFBFBF}51.03 \\
Llama3.3-70B-Instruct & Non-thinking & 70B & \cellcolor[HTML]{A6A6A6}62.95 & \cellcolor[HTML]{D2D2D2}34.18 \\
Qwen3-32B & Non-thinking & 32B & \cellcolor[HTML]{BABABA}45.59 & \cellcolor[HTML]{A6A6A6}72.82 \\
DeepSeeek-R1-0528 & Thinking & 671B & \cellcolor[HTML]{D4D4D4}21.84 & \cellcolor[HTML]{D1D1D1}35.11 \\
DeepSeek-V3.1-Terminus & Thinking & 671B & \cellcolor[HTML]{D3D3D3}22.58 & \cellcolor[HTML]{D2D2D2}34.18 \\
DeepSeek-V3.2 & Thinking & 671B & \cellcolor[HTML]{C9C9C9}31.90 & \cellcolor[HTML]{C4C4C4}46.64 \\
GLM-4.6-355B-A32B & Thinking & 355B & \cellcolor[HTML]{D8D8D8}17.89 & \cellcolor[HTML]{D6D6D6}30.68 \\
Qwen3-235B-A22B-Thinking-2507 & Thinking & 235B & \cellcolor[HTML]{FFFFFF}8.74 & \cellcolor[HTML]{F7F7F7}19.91 \\
Qwen3-32B & Thinking & 32B & \cellcolor[HTML]{D3D3D3}22.28 & \cellcolor[HTML]{FFFFFF}17.82\\ 
\bottomrule
\end{tabular}
}
\end{table}

Table \ref{tab:code-change} presents the unfaithful rates of each model across various reasoning modes and subsets. The results indicate that the models exhibit a higher unfaithfulness rate on \bench-2025, and thinking models are more faithful. However, no model achieved complete faithfulness in direct prompting. Despite such modifications typically not significantly impacting the execution results, they do introduce potential software engineering risks.

According to table \ref{tab:rq1}, after adding faithful-aware detection in the verification, the performance of the models decreases approximately 20\% on average. The following analysis are based on the faithful-aware pass@k (\ie code and asserts are both faithful and specifications pass the verify).

\subsubsection{Subsets} 
Table \ref{tab:rq1} indicates that all models generally exhibit inferior performance across all metrics on \bench-2025 compared to \bench-Pre2025, demonstrating that the newer dataset presents a greater challenge, which is compatible with statistical metrics presented in Section \ref{subsubsec:difficulty}.

\subsubsection{Metrics}
From a metric perspective, pass@$k$ steadily improves as $k$ increases for every model. Moreover, comparisons across different values of $k$ reveal consistency in the models' pass@k performance, which exhibits similar partial order relationships. This demonstrates that, under the direct-prompting setup, all models maintain a balance between stability (pass@k for small k) and exploration (pass@k for large k) across both subsets.

\subsubsection{Models}
Comparing the results of different models in Table \ref{tab:rq1}, we found thinking models outperform non-thinking models. This even holds true even for models with the same parameter weight, which will be further analyzed in the RQ2. In addition, the correlation between performance and model size is much more significant in non-thinking models. 

\newcounter{rq} 
\newcommand{\answerRQ}[1]{\refstepcounter{rq}
\vspace{1mm}
\begin{mdframed}[linecolor=gray,roundcorner=12pt,backgroundcolor=gray!15,linewidth=3pt,innerleftmargin=2pt, 
leftmargin=0cm, rightmargin=0cm, topline=false, bottomline=false, rightline=false]
\textbf{RQ\arabic{rq} Summary:} 
In direct prompting, models tend to exhibit non-negligible unfaithful behaviors by ignoring code context constraints. Filtering out these deceptive cases results in an approximate \textbf{20\% performance drop}. 

\noindent Increasing the sample size yields significant improvements: on average, pass@5 \textbf{doubles} pass@1, and pass@32 \textbf{triples} it, demonstrating the effectiveness of test-time scaling.
\end{mdframed}
\vspace{1mm}
}
\answerRQ{1}

\subsection{RQ2. Improvements from Thinking Mode}\label{sec:rq2}

\begin{table*}[tb!]
\centering
\caption{ \centering RQ2. How far can thinking mode improve the generation of formal specifications? \\
}
\label{tab:rq2}
\renewcommand\arraystretch{1.2}
    \resizebox{1.0\linewidth}{!}{
    \begin{tabular}{l|l|ll|l|ll|l|ll|l|ll|l|ll|l|ll|l}
    \toprule
 &  & \multicolumn{9}{c|}{\textbf{\bench-Pre2025}} & \multicolumn{9}{c}{\textbf{\bench-2025}} \\ \cline{3-20} 
 &  & \multicolumn{3}{c|}{Pass@1} & \multicolumn{3}{c|}{Pass@5} & \multicolumn{3}{c|}{Pass@32} & \multicolumn{3}{c|}{Pass@1} & \multicolumn{3}{c|}{Pass@5} & \multicolumn{3}{c}{Pass@32} \\ \cline{3-20} 
\multirow{-3}{*}{LLM} & \multirow{-3}{*}{Size} & \multicolumn{1}{c}{\faRobot~} & \multicolumn{1}{c}{\faLightbulb} & \multicolumn{1}{c|}{$\Uparrow$ (\%)} & \multicolumn{1}{c}{\faRobot~} & \multicolumn{1}{c}{\faLightbulb} & \multicolumn{1}{c|}{$\Uparrow$ (\%)} & \multicolumn{1}{c}{\faRobot~} & \multicolumn{1}{c}{\faLightbulb} & \multicolumn{1}{c|}{$\Uparrow$ (\%)} & \multicolumn{1}{c}{\faRobot~} & \multicolumn{1}{c}{\faLightbulb} & \multicolumn{1}{c|}{$\Uparrow$ (\%)} & \multicolumn{1}{c}{\faRobot~} & \multicolumn{1}{c}{\faLightbulb} & \multicolumn{1}{c|}{$\Uparrow$ (\%)} & \multicolumn{1}{c}{\faRobot~} & \multicolumn{1}{c}{\faLightbulb} & \multicolumn{1}{c}{$\Uparrow$ (\%)} \\
\midrule

DeepSeek-V3.1-Terminus & 671B & \cellcolor[HTML]{E3F5F2}24.64 & \cellcolor[HTML]{DCF3EF}30.52 & \cellcolor[HTML]{FFFBF4}23.86 & \cellcolor[HTML]{D3EFEB}38.30 & \cellcolor[HTML]{B7E5DE}62.71 & \cellcolor[HTML]{FFF4E2}63.73 & \cellcolor[HTML]{C4EAE4}51.11 & \cellcolor[HTML]{99DAD1}88.52 & \cellcolor[HTML]{FFF2DE}73.20 & \cellcolor[HTML]{FCFEFE}2.87 & \cellcolor[HTML]{F0FAF8}13.64 & \cellcolor[HTML]{FFDC9D}375.26 & \cellcolor[HTML]{F8FDFC}6.32 & \cellcolor[HTML]{E2F5F2}25.00 & \cellcolor[HTML]{FFDDA3}295.57 & \cellcolor[HTML]{F2FBF9}11.67 & \cellcolor[HTML]{D2EFEB}39.17 & \cellcolor[HTML]{FFDEA7}235.65 \\
DeepSeek-V3.2 & 671B & \cellcolor[HTML]{E5F6F3}23.24 & \cellcolor[HTML]{DCF3EF}30.31 & \cellcolor[HTML]{FFFAF1}30.42 & \cellcolor[HTML]{CEEDE9}43.08 & \cellcolor[HTML]{BAE6E0}59.93 & \cellcolor[HTML]{FFF8ED}39.11 & \cellcolor[HTML]{B9E6DF}60.74 & \cellcolor[HTML]{A0DDD4}82.22 & \cellcolor[HTML]{FFF9EF}35.36 & \cellcolor[HTML]{FDFFFE}2.13 & \cellcolor[HTML]{FAFEFD}4.65 & \cellcolor[HTML]{FFEAC9}118.31 & \cellcolor[HTML]{FAFEFD}4.76 & \cellcolor[HTML]{F0FAF9}13.09 & \cellcolor[HTML]{FFE0AF}175.00 & \cellcolor[HTML]{F6FCFB}8.33 & \cellcolor[HTML]{E2F5F2}25.83 & \cellcolor[HTML]{FFDEA9}210.08 \\
GLM-4.6-355B-A32B & 355B & \cellcolor[HTML]{ECF8F6}17.15 & \cellcolor[HTML]{E0F4F1}27.08 & \cellcolor[HTML]{FFF5E5}57.90 & \cellcolor[HTML]{D4F0EC}37.53 & \cellcolor[HTML]{B7E5DE}62.46 & \cellcolor[HTML]{FFF3E1}66.43 & \cellcolor[HTML]{BDE7E1}57.41 & \cellcolor[HTML]{96D9CF}91.11 & \cellcolor[HTML]{FFF5E4}58.70 & \cellcolor[HTML]{F7FCFC}7.37 & \cellcolor[HTML]{F1FAF9}12.30 & \cellcolor[HTML]{FFF3E1}66.89 & \cellcolor[HTML]{E8F7F5}20.65 & \cellcolor[HTML]{E0F4F1}27.42 & \cellcolor[HTML]{FFFAF1}32.78 & \cellcolor[HTML]{D6F0ED}35.83 & \cellcolor[HTML]{CEEEE9}42.78 & \cellcolor[HTML]{FFFCF7}19.40 \\
Qwen3-235B-A22B-Instruct-2507 & 235B & \cellcolor[HTML]{F2FBF9}11.54 & \cellcolor[HTML]{D7F1ED}34.92 & \cellcolor[HTML]{FFDEAA}202.60 & \cellcolor[HTML]{E7F7F4}21.35 & \cellcolor[HTML]{B8E6DF}61.51 & \cellcolor[HTML]{FFDEAB}188.10 & \cellcolor[HTML]{DBF2EF}31.11 & \cellcolor[HTML]{A4DED6}78.89 & \cellcolor[HTML]{FFE4B8}153.58 & \cellcolor[HTML]{FDFFFE}1.89 & \cellcolor[HTML]{F7FCFC}7.38 & \cellcolor[HTML]{FFDDA3}290.48 & \cellcolor[HTML]{F9FDFD}5.52 & \cellcolor[HTML]{E7F7F4}20.92 & \cellcolor[HTML]{FFDDA4}278.99 & \cellcolor[HTML]{F4FBFA}10.00 & \cellcolor[HTML]{D2EFEB}38.89 & \cellcolor[HTML]{FFDDA3}288.90 \\
Qwen3-32B & 32B & \cellcolor[HTML]{F8FDFC}6.33 & \cellcolor[HTML]{E0F4F1}27.44 & \cellcolor[HTML]{FFDDA0}333.49 & \cellcolor[HTML]{F0FAF8}13.77 & \cellcolor[HTML]{C3EAE4}52.16 & \cellcolor[HTML]{FFDDA4}278.79 & \cellcolor[HTML]{E2F5F2}25.19 & \cellcolor[HTML]{AAE1D9}73.33 & \cellcolor[HTML]{FFDEAB}191.11 & \cellcolor[HTML]{FFFFFF}0.29 & \cellcolor[HTML]{F7FCFC}7.44 & \cellcolor[HTML]{FFC000}2465.52 & \cellcolor[HTML]{FEFFFF}1.31 & \cellcolor[HTML]{E9F7F5}19.78 & \cellcolor[HTML]{FFCE50}1409.92 & \cellcolor[HTML]{F9FDFD}5.28 & \cellcolor[HTML]{D7F1ED}34.72 & \cellcolor[HTML]{FFDA8F}557.58

 \\
\bottomrule
\end{tabular}
}
\end{table*}

\tablename~\ref{tab:rq2} presents the effect of enabling \textit{thinking model} versus \textit{non-thinking mode} on formal specification generation across the two \bench subsets. 

Overall, thinking mode yields consistent and substantial improvements for all models and all metrics (pass@1/5/32), with relative gains ranging from 19.40\% to 2465.52\%. 
This indicates that making reasoning more explicit helps address the intrinsic difficulty of mapping natural-language requirements into formal specifications, improving both first-attempt success (pass@1) and multi-attempt success (pass@5 and pass@32) during inference.

Thinking mode tends to bring larger benefits on the more unfamiliar or newer subset (\bench-2025).
On \bench-Pre2025, the relative improvements in pass@1 span from 23.86\% to 333.49\%, with a median gain of 57.90\%.
In contrast, on \bench-2025, the range shifts to 66.89\% \textasciitilde 2465.52\%, and the median gain increases to 290.48\%.
This phenomenon suggests that when facing more distribution-shifted or new problems, the thinking mode better leverages step-by-step reasoning and self-checking, leading to large relative gains.

From the perspective of model scale, we observe a trend that smaller models benefit more from the thinking mode. 
For instance, on \bench-Pre2025, Qwen3-32B increases its pass@5 from 6.33 to 27.44 (+21.11 absolute points, +333.49\%).
On \bench-2025, the same model’s pass@5 rises from 0.29 to 7.44 (+7.15 absolute points, +2465.52\%, over 25x).

\refstepcounter{rq}
\vspace{1mm}
\begin{mdframed}[linecolor=gray,roundcorner=12pt,backgroundcolor=gray!15,linewidth=3pt,innerleftmargin=2pt, 
leftmargin=0cm, rightmargin=0cm, topline=false, bottomline=false, rightline=false]
\textbf{RQ\arabic{rq} Summary:} 
The thinking mode yields \textbf{substantial improvements}, with relative
gains ranging from 19.40\% to 2465.52\%. The magnitude of these improvements grows \textbf{as dataset difficulty increases or model size decreases}. Also, \textbf{smaller models} tend to benefit more from the thinking mode (e.g., Qwen3-32B pass@5 raises from 6.33 to 27.44). 
\end{mdframed}
\vspace{1mm}

\subsection{RQ3. Improvements from Agentic Pipeline}\label{sec:rq3}

In this subsection, we evaluate AutoSpec \cite{wen2024enchantingprogramspecificationsynthesis} and compare it against direct prompting with the same base models.

Table \ref{tab:rq3} presents the results of the AutoSpec workflow. Compared to direct prompting, AutoSpec yields an unstable performance gain. Most models perform better in AutoSpec than in direct prompting. However, one notable exception is R1, which exhibits a performance decline across all three metrics, with the drop reaching as much as 50\% on \bench-2025. 

The growth from pass@1 to pass@5 under the AutoSpec pipeline is more marginal than the direct prompting, exhibiting higher stability across different samples. This is probably attributed to the integration of fixed processing modules of codes and specs within AutoSpec, which leads to a standardized response.

We further perform an ablation study by decomposing the agent into its constituent components. Each iteration of AutoSpec primarily relies on two modules: the program analysis module and the verifier-as-feedback module. We observe that relying solely on the program analysis module consistently degrades performance, with decreases ranging from 15\% to 75\%, demonstrating the critical role of the verifier feedback module. However, when employing only the verifier, the pass@1 metric decreases solely for non-thinking models on the \bench-Pre2025 subset; performance across all other configurations improves. The performance gain when relying solely on verifier feedback reaches as high as 255.99\%, over 3x. This is likely because, while the current program analysis module assists the model in decomposing and solving the target program, it simultaneously causes the model to overlook the underlying connections between program components. A case analysis of AutoSpec's errors is provided in Listing \ref{lst:failure-autospec} in the subsequent section.

In addition, iteratively generating specifications for 5 rounds yields significant gains compared to performing 5 independent sampling attempts from scratch.Taking DeepSeek-V3.2 as an example, on \bench-Pre2025, the accuracies for 5 independent sampling attempts and 5 rounds of iterative sampling are 67.04\% and 79.26\%, respectively, representing an increase of approximately 18\%. On \bench-2025, the corresponding accuracies are 25.83\% and 57.50\%, reflecting an improvement of over 100\%. This indicates that the agent is capable of leveraging verifier feedback to a certain extent, especially in harder datasets.

\begin{table*}[tb!]
\caption{RQ3. How Much Can the Agent Pipeline Improve Formal Specification Generation?
}
\label{tab:rq3}
\renewcommand\arraystretch{1.2}
 \resizebox{1.0\linewidth}{!}{
\begin{tabular}{lccccccccccccccc}
\toprule
\multicolumn{16}{c}{\textbf{\bench-pre2025}} \\ \hline
\multicolumn{1}{l|}{\textbf{}} & \multicolumn{7}{c|}{\textbf{pass@1}} & \multicolumn{7}{c|}{\textbf{pass@5}} & {\textbf{Iter@5}} \\ \cline{2-16} 
\multicolumn{1}{c|}{} & \multicolumn{1}{c|}{} & \multicolumn{2}{c|}{\textbf{End to end}} & \multicolumn{2}{c|}{\textbf{Program Analysis Only}} & \multicolumn{2}{c|}{\textbf{Verifier-as-feedback Only}} & \multicolumn{1}{c|}{} & \multicolumn{2}{c|}{\textbf{End to end}} & \multicolumn{2}{c|}{\textbf{Program Analysis Only}} & \multicolumn{2}{c|}{\textbf{Verifier-as-feedback Only}} &  \\ \cline{3-8} \cline{10-15}
\multicolumn{1}{c|}{\multirow{-2}{*}{\textbf{LLMs}}} & \multicolumn{1}{c|}{\multirow{-2}{*}{\textbf{Agent}}} & \textbf{Value} & \multicolumn{1}{c|}{$\Downarrow$ (\%)} & \textbf{Value} & \multicolumn{1}{c|}{$\Downarrow$ (\%)} & \textbf{Value} & \multicolumn{1}{c|}{$\Downarrow$ (\%)} & \multicolumn{1}{c|}{\multirow{-2}{*}{\textbf{Agent}}} & \textbf{Value} & \multicolumn{1}{c|}{$\Downarrow$ (\%)} & \textbf{Value} & \multicolumn{1}{c|}{$\Downarrow$ (\%)} & \textbf{Value} & \multicolumn{1}{c|}{$\Downarrow$ (\%)} & \multirow{-2}{*}{\textbf{Agent}} \\ 

\midrule

\multicolumn{1}{l|}{\faRobot~Llama3.3-70B-Instruct} & \multicolumn{1}{c|}{\cellcolor[HTML]{C0E9E3}51.48} & \cellcolor[HTML]{FAFEFD}4.27 & \multicolumn{1}{c|}{\cellcolor[HTML]{BFBFBF}91.71} & \cellcolor[HTML]{D8F1EE}31.93 & \multicolumn{1}{c|}{\cellcolor[HTML]{DDDDDF}37.98} & \cellcolor[HTML]{CAECE7}43.77 & \multicolumn{1}{c|}{\cellcolor[HTML]{F0F0F3}14.98} & \multicolumn{1}{c|}{\cellcolor[HTML]{B9E6DF}57.41} & \cellcolor[HTML]{F2FBF9}10.83 & \multicolumn{1}{c|}{\cellcolor[HTML]{BFBFBF}81.14} & \cellcolor[HTML]{CEEEE9}40 & \multicolumn{1}{c|}{\cellcolor[HTML]{E4E4E5}30.33} & \cellcolor[HTML]{ADE1DA}67.38 & \multicolumn{1}{c|}{\cellcolor[HTML]{FDF9F9}-17.37} & \cellcolor[HTML]{99DAD0}70.37 \\
\multicolumn{1}{l|}{\faRobot~Qwen3-32B} & \multicolumn{1}{c|}{\cellcolor[HTML]{C2E9E4}49.63} & \cellcolor[HTML]{F8FDFC}6.33 & \multicolumn{1}{c|}{\cellcolor[HTML]{C4C4C5}87.25} & \cellcolor[HTML]{D8F1ED}32.15 & \multicolumn{1}{c|}{\cellcolor[HTML]{E0E0E1}35.22} & \cellcolor[HTML]{D2EFEB}36.59 & \multicolumn{1}{c|}{\cellcolor[HTML]{E7E7E9}26.27} & \multicolumn{1}{c|}{\cellcolor[HTML]{B8E6DF}58.15} & \cellcolor[HTML]{EFF9F8}13.77 & \multicolumn{1}{c|}{\cellcolor[HTML]{C3C3C3}76.32} & \cellcolor[HTML]{C9ECE7}44.44 & \multicolumn{1}{c|}{\cellcolor[HTML]{E9E9EB}23.58} & \cellcolor[HTML]{B5E5DE}60.15 & \multicolumn{1}{c|}{\cellcolor[HTML]{FDFBFD}-3.44} & \cellcolor[HTML]{A0DDD4}65.56 \\
\multicolumn{1}{l|}{\faLightbulb~Qwen3-32B} & \multicolumn{1}{c|}{\cellcolor[HTML]{C0E9E3}51.56} & \cellcolor[HTML]{DEF3F0}27.44 & \multicolumn{1}{c|}{\cellcolor[HTML]{F2F2F4}46.78} & \cellcolor[HTML]{D6F1ED}33.48 & \multicolumn{1}{c|}{\cellcolor[HTML]{E0E0E1}35.07} & \cellcolor[HTML]{BCE7E1}55.07 & \multicolumn{1}{c|}{\cellcolor[HTML]{FDFBFC}-6.81} & \multicolumn{1}{c|}{\cellcolor[HTML]{B0E3DB}64.44} & \cellcolor[HTML]{BFE8E2}52.16 & \multicolumn{1}{c|}{\cellcolor[HTML]{EEEEF0}19.06} & \cellcolor[HTML]{C0E8E2}51.85 & \multicolumn{1}{c|}{\cellcolor[HTML]{ECECEF}19.54} & \cellcolor[HTML]{9CDBD2}80.93 & \multicolumn{1}{c|}{\cellcolor[HTML]{FDF9F6}-25.59} & \cellcolor[HTML]{9EDCD3}66.3 \\
\multicolumn{1}{l|}{\faLightbulb~DeepSeeek-R1-0528} & \multicolumn{1}{c|}{\cellcolor[HTML]{BCE7E1}55.11} & \cellcolor[HTML]{D5F0EC}34.69 & \multicolumn{1}{c|}{\cellcolor[HTML]{FCFCFF}37.05} & \cellcolor[HTML]{EEF9F8}14.22 & \multicolumn{1}{c|}{\cellcolor[HTML]{BFBFBF}74.20} & \cellcolor[HTML]{AEE2DA}66.02 & \multicolumn{1}{c|}{\cellcolor[HTML]{FDF9F8}-19.80} & \multicolumn{1}{c|}{\cellcolor[HTML]{AFE2DB}65.56} & \cellcolor[HTML]{AAE0D8}69.63 & \multicolumn{1}{c|}{\cellcolor[HTML]{FFDEAB}-6.21} & \cellcolor[HTML]{D4F0EC}35.56 & \multicolumn{1}{c|}{\cellcolor[HTML]{D7D7D8}45.76} & \cellcolor[HTML]{90D7CD}90.4 & \multicolumn{1}{c|}{\cellcolor[HTML]{FDF7F2}-37.89} & \cellcolor[HTML]{97DAD0}71.11 \\
\multicolumn{1}{l|}{\faLightbulb~DeepSeeek-v3.2} & \multicolumn{1}{c|}{\cellcolor[HTML]{B5E4DD}60.74} & \cellcolor[HTML]{DAF2EE}30.31 & \multicolumn{1}{c|}{\cellcolor[HTML]{EEEEF0}50.10} & \cellcolor[HTML]{DAF2EF}30.07 & \multicolumn{1}{c|}{\cellcolor[HTML]{D3D3D4}50.49} & \cellcolor[HTML]{A3DED5}75.53 & \multicolumn{1}{c|}{\cellcolor[HTML]{FDF9F7}-24.35} & \multicolumn{1}{c|}{\cellcolor[HTML]{ADE2DA}67.04} & \cellcolor[HTML]{B6E5DE}59.93 & \multicolumn{1}{c|}{\cellcolor[HTML]{F5F5F7}10.61} & \cellcolor[HTML]{BAE6E0}56.67 & \multicolumn{1}{c|}{\cellcolor[HTML]{F0F0F2}15.47} & \cellcolor[HTML]{8BD5CA}94.24 & \multicolumn{1}{c|}{\cellcolor[HTML]{FDF7F1}-40.57} & \cellcolor[HTML]{8BD5CA}79.26 \\ 

\midrule

\multicolumn{16}{c}{\textbf{\bench-2025}} \\ \hline
\multicolumn{1}{l|}{} & \multicolumn{7}{c|}{\textbf{pass@1}} & \multicolumn{7}{c|}{\textbf{pass@5}} & {\textbf{Iter@5}} \\ \cline{2-16} 
\multicolumn{1}{c|}{} & \multicolumn{1}{c|}{} & \multicolumn{2}{c|}{\textbf{End to end}} & \multicolumn{2}{c|}{\textbf{Program Analysis Only}} & \multicolumn{2}{c|}{\textbf{Verifier-as-feedback Only}} & \multicolumn{1}{c|}{} & \multicolumn{2}{c|}{\textbf{End to end}} & \multicolumn{2}{c|}{\textbf{Program Analysis Only}} & \multicolumn{2}{c|}{\textbf{Verifier-as-feedback Only}} &  \\ \cline{3-8} \cline{10-15}
\multicolumn{1}{c|}{\multirow{-2}{*}{\textbf{LLMs}}} & \multicolumn{1}{c|}{\multirow{-2}{*}{\textbf{Agent}}} & \textbf{Value} & \multicolumn{1}{c|}{$\Downarrow$ (\%)} & \textbf{Value} & \multicolumn{1}{c|}{$\Downarrow$ (\%)} & \textbf{Value} & \multicolumn{1}{c|}{$\Downarrow$ (\%)} & \multicolumn{1}{c|}{\multirow{-2}{*}{\textbf{Agent}}} & \textbf{Value} & \multicolumn{1}{c|}{$\Downarrow$ (\%)} & \textbf{Value} & \multicolumn{1}{c|}{$\Downarrow$ (\%)} & \textbf{Value} & \multicolumn{1}{c|}{$\Downarrow$ (\%)} & \multirow{-2}{*}{\textbf{Agent}} \\ \hline
\multicolumn{1}{l|}{\faRobot~Llama3.3-70B-Instruct} & \multicolumn{1}{c|}{\cellcolor[HTML]{EAF8F5}17.83} & \cellcolor[HTML]{FFFFFF}0.25 & \multicolumn{1}{c|}{\cellcolor[HTML]{BFBFBF}98.60} & \cellcolor[HTML]{F4FBFA}9.61 & \multicolumn{1}{c|}{\cellcolor[HTML]{D7D7D8}46.10} & \cellcolor[HTML]{E5F6F4}21.28 & \multicolumn{1}{c|}{\cellcolor[HTML]{FDF9F8}-19.35} & \multicolumn{1}{c|}{\cellcolor[HTML]{E5F6F3}21.67} & \cellcolor[HTML]{FEFFFF}0.82 & \multicolumn{1}{c|}{\cellcolor[HTML]{BFBFBF}96.22} & \cellcolor[HTML]{EEF9F7}14.44 & \multicolumn{1}{c|}{\cellcolor[HTML]{E1E1E3}33.36} & \cellcolor[HTML]{C3EAE4}48.88 & \multicolumn{1}{c|}{\cellcolor[HTML]{FEEDD5}-125.57} & \cellcolor[HTML]{CBECE7}36.11 \\
\multicolumn{1}{l|}{\faRobot~Qwen3-32B} & \multicolumn{1}{c|}{\cellcolor[HTML]{ECF9F7}15.56} & \cellcolor[HTML]{FFFFFF}0.29 & \multicolumn{1}{c|}{\cellcolor[HTML]{C0C0C0}98.14} & \cellcolor[HTML]{F3FBFA}9.78 & \multicolumn{1}{c|}{\cellcolor[HTML]{DEDEDF}37.15} & \cellcolor[HTML]{DBF2EF}29.77 & \multicolumn{1}{c|}{\cellcolor[HTML]{FEF1E1}-91.32} & \multicolumn{1}{c|}{\cellcolor[HTML]{E4F6F3}22.22} & \cellcolor[HTML]{FEFFFF}1.31 & \multicolumn{1}{c|}{\cellcolor[HTML]{C1C1C1}94.10} & \cellcolor[HTML]{ECF9F7}15.56 & \multicolumn{1}{c|}{\cellcolor[HTML]{E4E4E6}29.97} & \cellcolor[HTML]{C1E9E3}50.39 & \multicolumn{1}{c|}{\cellcolor[HTML]{FEEDD5}-126.78} & \cellcolor[HTML]{D1EFEA}31.67 \\
\multicolumn{1}{l|}{\faLightbulb~Qwen3-32B} & \multicolumn{1}{c|}{\cellcolor[HTML]{EBF8F6}16.5} & \cellcolor[HTML]{F6FCFB}7.44 & \multicolumn{1}{c|}{\cellcolor[HTML]{DBDBDC}54.91} & \cellcolor[HTML]{F5FCFB}8.72 & \multicolumn{1}{c|}{\cellcolor[HTML]{D6D6D7}47.15} & \cellcolor[HTML]{D9F2EE}31.09 & \multicolumn{1}{c|}{\cellcolor[HTML]{FEF1E1}-88.42} & \multicolumn{1}{c|}{\cellcolor[HTML]{E2F5F2}23.89} & \cellcolor[HTML]{E7F7F4}19.78 & \multicolumn{1}{c|}{\cellcolor[HTML]{F2F2F4}17.20} & \cellcolor[HTML]{EAF8F6}17.22 & \multicolumn{1}{c|}{\cellcolor[HTML]{E6E6E7}27.92} & \cellcolor[HTML]{AAE0D8}69.6 & \multicolumn{1}{c|}{\cellcolor[HTML]{FFE5C0}-191.34} & \cellcolor[HTML]{CDEDE8}34.44 \\
\multicolumn{1}{l|}{\faLightbulb~DeepSeeek-R1-0528} & \multicolumn{1}{c|}{\cellcolor[HTML]{EEF9F7}14.44} & \cellcolor[HTML]{ECF8F7}15.73 & \multicolumn{1}{c|}{\cellcolor[HTML]{FFDEAB}-8.93} & \cellcolor[HTML]{F8FDFC}5.72 & \multicolumn{1}{c|}{\cellcolor[HTML]{CBCBCB}60.39} & \cellcolor[HTML]{C7EBE6}45.78 & \multicolumn{1}{c|}{\cellcolor[HTML]{FFE2B7}-217.04} & \multicolumn{1}{c|}{\cellcolor[HTML]{E6F6F4}21.11} & \cellcolor[HTML]{D8F1EE}31.81 & \multicolumn{1}{c|}{\cellcolor[HTML]{FFDEAB}-50.69} & \cellcolor[HTML]{F3FBFA}10.56 & \multicolumn{1}{c|}{\cellcolor[HTML]{D3D3D4}49.98} & \cellcolor[HTML]{A3DED5}75.15 & \multicolumn{1}{c|}{\cellcolor[HTML]{FFDEAB}-255.99} & \cellcolor[HTML]{C1E9E3}42.5 \\
\multicolumn{1}{l|}{\faLightbulb~DeepSeeek-v3.2} & \multicolumn{1}{c|}{\cellcolor[HTML]{E4F5F3}22.56} & \cellcolor[HTML]{FAFDFD}4.65 & \multicolumn{1}{c|}{\cellcolor[HTML]{CBCBCC}79.39} & \cellcolor[HTML]{F7FCFB}7.28 & \multicolumn{1}{c|}{\cellcolor[HTML]{C5C5C5}67.73} & \cellcolor[HTML]{DCF3EF}28.49 & \multicolumn{1}{c|}{\cellcolor[HTML]{FDF8F6}-26.29} & \multicolumn{1}{c|}{\cellcolor[HTML]{E0F4F1}25.83} & \cellcolor[HTML]{EFFAF8}13.09 & \multicolumn{1}{c|}{\cellcolor[HTML]{DDDDDF}49.32} & \cellcolor[HTML]{E8F7F5}19.44 & \multicolumn{1}{c|}{\cellcolor[HTML]{E8E8EA}24.74} & \cellcolor[HTML]{BBE7E0}55.27 & \multicolumn{1}{c|}{\cellcolor[HTML]{FEEED9}-113.98} & \cellcolor[HTML]{ABE1D9}57.5 \\ 
\bottomrule
\end{tabular}
}
\end{table*}

\refstepcounter{rq}
\vspace{1mm}
\begin{mdframed}[linecolor=gray,roundcorner=12pt,backgroundcolor=gray!15,linewidth=3pt,innerleftmargin=2pt, 
leftmargin=0cm, rightmargin=0cm, topline=false, bottomline=false, rightline=false]
\textbf{RQ\arabic{rq} Summary:} 
The agentic pipeline effectively improves the quality of generated specifications under a small number of sampling attempts (e.g., $k \leq 5$). These performance gains are more pronounced at \textbf{lower sampling budgets} and on \textbf{more challenging datasets}. However, as sampling attempts increase, \textbf{the improvements gradually diminish} to a marginal extent. Regarding the agent's components, we observe that the program analysis module is effective solely for non-thinking models, whereas both verifier feedback and iterative refinement yield substantial improvements exceeding 100\%. 

\end{mdframed}
\vspace{1mm}

\subsection{RQ4. Failure Analysis}
\definecolor{color-green}{HTML}{E6F1EA}
\definecolor{color-red}{HTML}{faccd0}
\definecolor{color-yellow}{HTML}{FCF3D5}

\subsubsection{Case Study}
\label{subsubsec:case-study}
We present several cases of errors spanning two approaches, stemming from diverse failure causes. Errors (unfaithful behavior or failed specifications) are highlighted in \colorbox{color-red}{red}, with the corresponding correct versions displayed in \colorbox{color-green}{green}.
\colorlet{FancyVerbHighlightColor}{color-red}
\begin{listing}[htbp]
\begin{minted}{cpp}
#define N 100000 
|\xglobal\colorlet{FancyVerbHighlightColor}{color-green}|int N = 100000
// |\textbf{\textcolor{black}{In source code, N should be a global variable rather than macro define}}|
int main(){
  int a[N];
  int b[N];
  int sum=0;
  // ... [Codes and specs are omitted for brevity] ...
  //@ assert sum == 0;
  return 0;
}
\end{minted}
\caption{A case of model infaithful generating code}
\label{lst:failure-faith-code}
\end{listing}

Listing \ref{lst:failure-faith-code} illustrates that the model may alter the original code during generation. For instance, the model converted a global variable defined in the input into a macro definition. This behavior likely stems from coding habits developed during the model's training phase. Although this specific modification did not interfere with the execution results or the verification goal within this file, altering variable definitions in such a manner can introduce security risks in a software engineering context.

\colorlet{FancyVerbHighlightColor}{color-red}
\begin{listing}[htbp]
\begin{minted}{cpp}
#include <limits.h>
#include <stdlib.h>
/*@ assigns and ensures are omitted here*/
int unknown();
const int MAX = 100000;
int main(){
  int SIZE = unknown();
  if (SIZE > 1 && SIZE < MAX){
    int i;
    int* a = malloc(sizeof(int) * SIZE);
    // ... [Codes and specs are omitted for brevity] ...
    for (i = 0; i < SIZE; i++){
       /*@ assert a[i] == (i >> 16 > 250 ? 1 : 0); */
       |\xglobal\colorlet{FancyVerbHighlightColor}{color-green}|//@ assert a[i] == 0
      // |\textbf{\textcolor{black}{LLM weakened the assert expression, trying to pass the evaluation}} |
    }
  }
  return 1;
}
\end{minted}

\caption{A case of model infaithful generating specification}
\label{lst:failure-faith-spec}
\end{listing}

Listing \ref{lst:failure-faith-spec} demonstrates how a model might \textit{cheat} during the direct prompting by tampering with input assertions. The original assertion required proving that all elements of array a are zero; however, the model-generated assertion allows for the existence of the value 1 within the array. The model failed to grasp that within the given range of i, a[i] could not possibly be assigned 1 in the first loop; instead, it simply modified the assertion to force a successful verification. While both examples pass the verifier, the resulting code-specification pairs fail to meet the rigorous requirements of formal verification. This highlights the necessity of our faithfulness-aware evaluation design.

\colorlet{FancyVerbHighlightColor}{color-red}
\begin{listing}[htbp]
\begin{minted}{cpp}
#include<limits.h>
/*@ assigns and ensures are omitted here*/
int unknown();
#include<stdlib.h>
int SIZE;
const int MAX = 100000;
int main(){
  SIZE = unknown();
  if(SIZE > 1 && SIZE < MAX) {
    long long i;
    long long *a = malloc(sizeof(long long)*SIZE);
    // ... [Codes and specs are omitted for brevity] ...
    |\xglobal\colorlet{FancyVerbHighlightColor}{color-green}|/*@
        loop invariant i <= SIZE;
        loop invariant 0 <= i;
        loop assigns i;
    */
    // |\textbf{\textcolor{black}{Under the end-to-end inference mode, the model omitted the specification for this loop}}|
	for(i=0; i<SIZE; i++)	{
	   //@ assert(a[i] == i*i);
	}
  }
  return 1;
}
\end{minted}

\caption{A case of a thinking model failing to prove the goal assertion in direct-prompting manner}
\label{lst:failure-endtoend}
\end{listing}
The remaining two examples illustrate specifications that failed verification. For the direct-prompting mode, we observed that the model occasionally misses specific locations where specifications should be written during the "thinking" process; this is likely due to the absence of a dedicated specification localization module. For instance, in Listing \ref{lst:failure-endtoend}, the model overlooks the invariant and variant specifications for the third loop. Without utilizing specifications to refine the value range of the loop variable $i$, the assertions within the loop cannot be verified by Frama-C.
\colorlet{FancyVerbHighlightColor}{color-red}
\begin{listing}[htbp]
\begin{minted}{cpp}
#include<limits.h>
/*@ assigns and ensures are omitted here*/
int unknown();
int CELLCOUNT;
int main(){
CELLCOUNT = unknown();
if(CELLCOUNT > 1)  {
    // ... [Codes and specs are omitted for brevity] ...
    int volArray[CELLCOUNT];
    if(CELLCOUNT % 2 != 0) { return 1; }
    //@ assert(CELLCOUNT % 2 == 0);
    /*@ loop invariant 1 <= i; */
    $\xglobal\colorlet{FancyVerbHighlightColor}{color-green}$/*@ 
        loop invariant 1 <= i <= CELLCOUNT/2 + 1;
        loop invariant \forall int k; 0 <= k < (i-1)*2 ==> volArray[k] >= MINVAL || volArray[k] == 0;
        loop assigns i, j, volArray[0 .. CELLCOUNT-1];
    */
    // $\textbf{\textcolor{black}{Within the inference framework of the autospec workflow, the model neglected the outer loop invariants pertaining to the inner loop}}$
    for(i = 1; i <= CELLCOUNT/2; i++)
        // ... [Specs are omitted for brevity] ...
        for(j = 2; j >= 1; j--){
            if(j >= MINVAL){
                volArray[i*2 - j] = j;
            }
            else{
                volArray[i*2 - j] = 0;
            }
        }
    for(i = 0; i < CELLCOUNT; i++){
        //@ assert(volArray[i] >= MINVAL || volArray[i] == 0 );
    }
  }
  return 1;
}
\end{minted}

\caption{A case where the thinking model fails to prove the property in agentic workflow}
\label{lst:failure-autospec}
\end{listing}
Regarding the AutoSpec inference mode, we found that the model significantly omits specifications for outer loops in nested loop structures. As shown in Listing \ref{lst:failure-autospec}, the model is expected to eliminate the inner loop variables within the outer loop invariants by leveraging the inner invariants to describe the memory state of volArray modified during the cycle.

\subsubsection{Failure Taxonomy}
We conducted a failure taxonomy on the model-generated specifications shared by both the \textit{direct prompting} and \textit{agentic pipeline}. Each file contains three categories of errors: typo, missing, and extra, which respectively represent specifications that are slightly misinterpreted, omitted, and spuriously generated by the model in incorrect outputs. Specifically, we use \texttt{difflib}~\cite{xu2025difflib} to compute the text similarity between each specification in the erroneous output and each specification in the corresponding ground truth. If a pair exhibits a similarity score between 0.8 and 0.99, it is classified as a typo and subsequently removed. For those that cannot be paired and removed, each remaining specification in the ground truth is counted as a missing error, while each remaining specification in model's output is counted as an extra error. Finally, this paper averages the error statistics across all files and organizes them according to their corresponding specification types.

\begin{figure}[tbp] 
     \centering
     \begin{subfigure}[b]{1.0\linewidth}
         \centering
         \includegraphics[width=\textwidth]{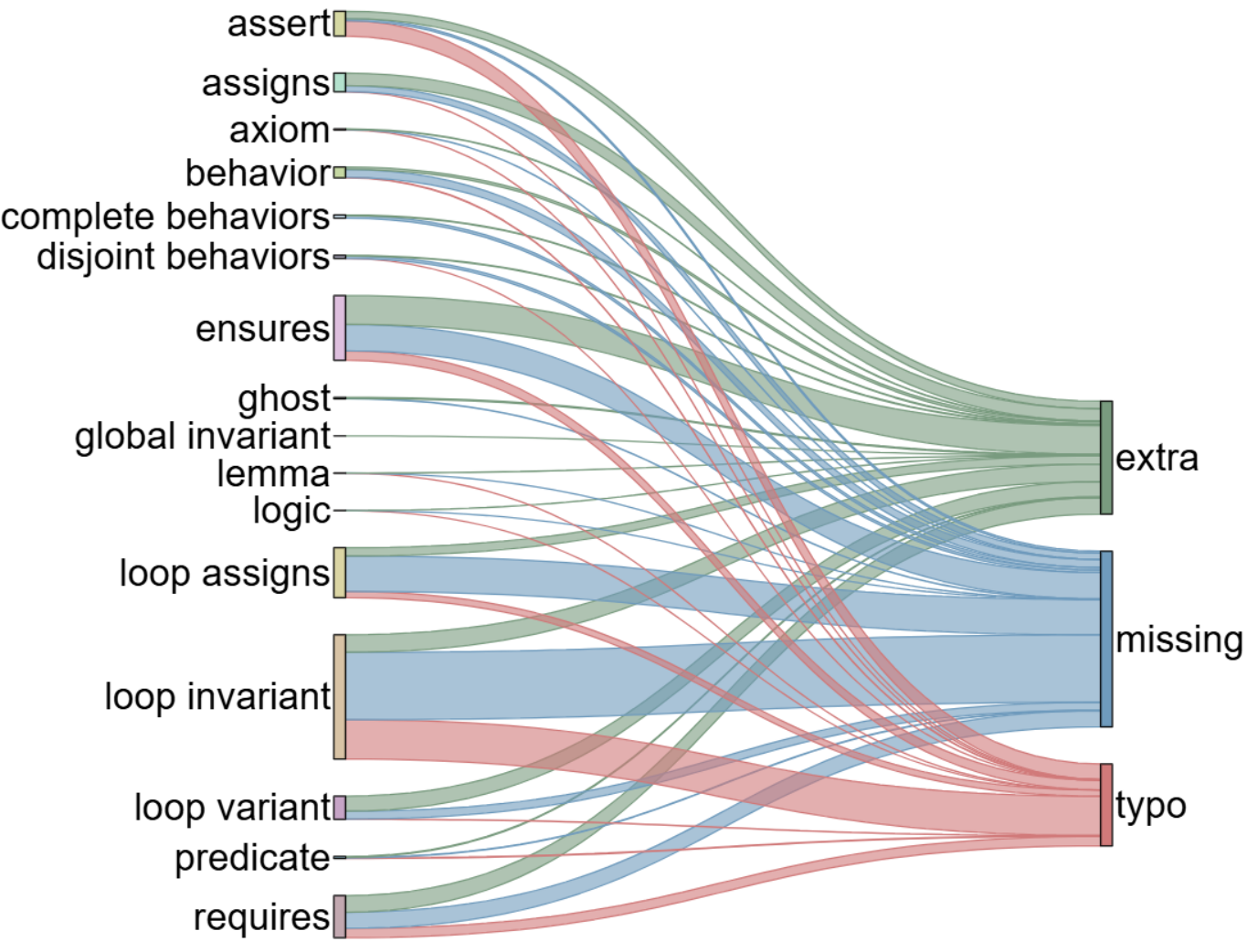}
         \caption{\centering Direct prompting  
         }
         \label{fig:end2end taxonomy}
     \end{subfigure}
     \hfill 
     \begin{subfigure}[b]{1.0\linewidth}
         \centering
         \includegraphics[width=\textwidth]{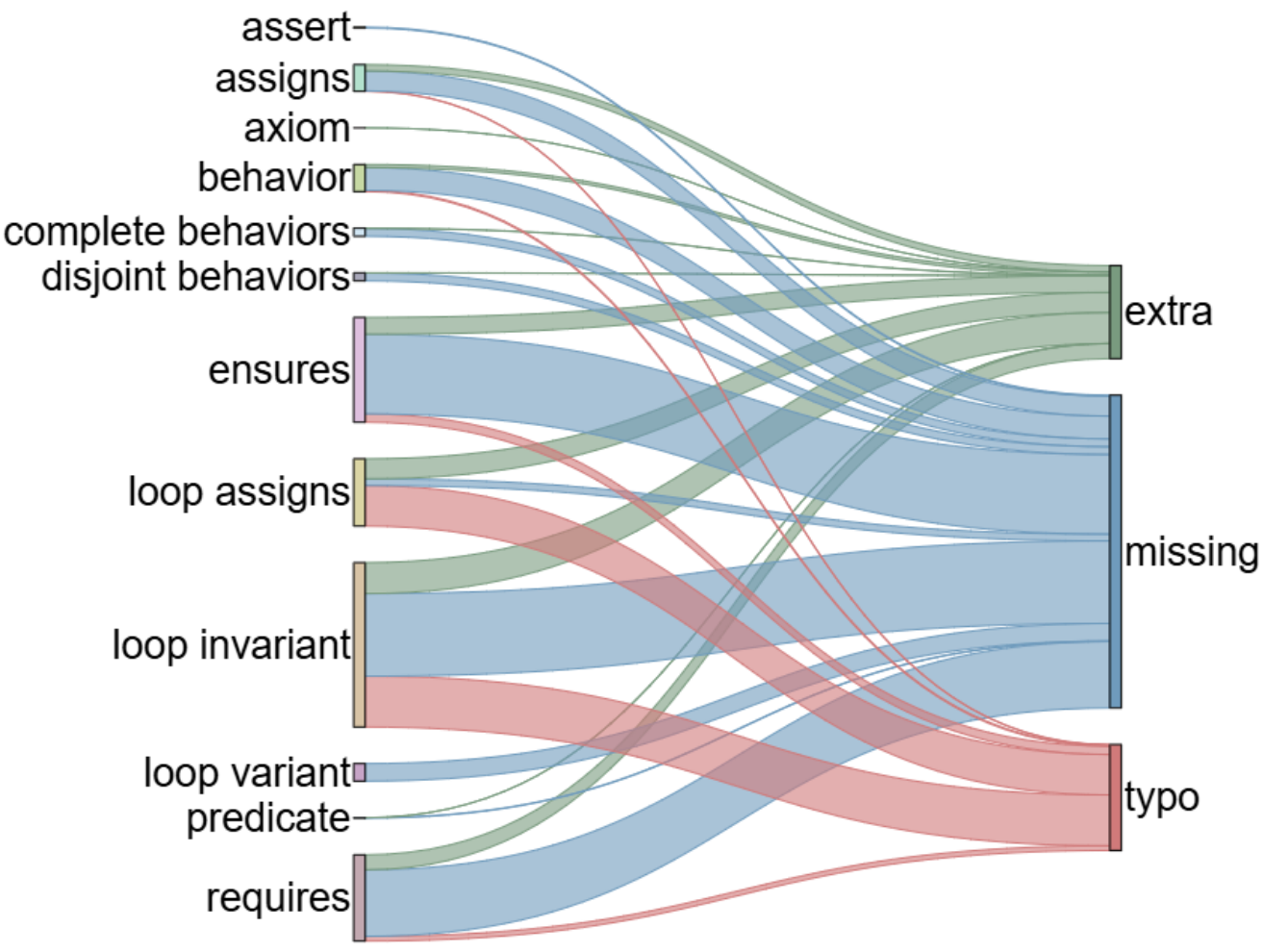}
         \caption{\centering Agentic pipeline}
         \label{fig:autospec taxonomy}
     \end{subfigure}
     
     \caption{ Aggregated Failure taxonomy }
     \label{fig:failure taxonomy}
\end{figure}

In terms of \textbf{error types}, loop invariant errors account for the highest proportion across both reasoning modes, which is shown in Figure \ref{fig:failure taxonomy}. \texttt{Pre-conditions}, \texttt{post-conditions}, and \texttt{loop assigns} are also specification types with relatively concentrated error distributions. Assertion errors are significantly less frequent in the agentic pipeline. This is because the model may improperly alter assertions within the direct-prompting pipeline.

As for the error categories, compared to the ground truth, omitted specifications are more common than redundant ones. Notably, the agentic pipeline produces fewer redundant specifications than the direct-prompting mode. This is likely because the task decomposition in the agentic pipeline makes models focus on variable relationships within a specific module. This difference is most pronounced in ensure and require specifications (post- and pre-conditions), as the direct-prompting pipeline prompts the model to focus more easily on how a function interacts with other modules in the global scope.

\refstepcounter{rq}
\vspace{1mm}
\begin{mdframed}[linecolor=gray,roundcorner=12pt,backgroundcolor=gray!15,linewidth=3pt,innerleftmargin=2pt, 
leftmargin=0cm, rightmargin=0cm, topline=false, bottomline=false, rightline=false]
\textbf{RQ\arabic{rq} Summary:} 
The failure analysis shows that \textbf{incorrect loop invariant} account for the highest proportion across both reasoning modes. Agentic pipeline significantly reduces the specifications that violate the assertion to be verified (i.e., assertion error)
\end{mdframed}
\vspace{1mm}

\subsection{RQ5: Inference Cost}
Among the three test-time scaling approaches, increasing the number of generation attempts ($k$ in $pass@k$) inevitably leads to a proportional increase in token consumption. The results from \ref{sec:rq1} demonstrate that this method enhance model performance.

We then measured token consumption across various inference modes to assess the cost of generating specifications for other two scaling methods, as shown in Table \ref{tab:cost}. Compared to non-thinking models, thinking models consume significantly more tokens and exhibit greater sensitivity to the difficulty of the input data, for the token consumption of reasoning models is markedly higher on the more challenging \bench-2025 (1957.45 vs. 2414.12 for min and 4383.45 vs. 5587.93 for max). Additionally, AutoSpec incurs higher token consumption than direct prompting, especially for non-thinking models. Both the explicit reasoning process and the Agent workflow trade higher inference overhead for more accurate specification generation.

\begin{table}[htbp]
\caption{RQ5. Token cost and pass rate comparison}
\resizebox{1.00\linewidth}{!}{
\begin{tabular}{l|cccc}
\toprule
\multicolumn{1}{c|}{} & \multicolumn{4}{c}{\textbf{\bench-Pre2025}} \\ \cline{2-5} 
\multicolumn{1}{c|}{} & \multicolumn{2}{c|}{\textbf{End to End}} & \multicolumn{2}{c}{\textbf{AutoSpec}} \\
\multicolumn{1}{c|}{\multirow{-3}{*}{\textbf{LLMs}}} & {\textbf{pass@1}} & \multicolumn{1}{c|}{\textbf{\#Token}} & {\textbf{pass@1}} & \textbf{\#Token}  \\ \hline
\faRobot~Llama3.3-70B-Instruct & \cellcolor[HTML]{F7FDFC}4.27 & \multicolumn{1}{c|}{\cellcolor[HTML]{FCFCFF}221.02} & \cellcolor[HTML]{9CDCD2}52.22 & \cellcolor[HTML]{FCFCFF}2009.65 \\
\faRobot~Qwen3-32B & \cellcolor[HTML]{F3FBFA}6.33 & \multicolumn{1}{c|}{\cellcolor[HTML]{FCFCFF}207.77} & \cellcolor[HTML]{A1DDD4}49.70 & \cellcolor[HTML]{FCFCFF}2001.10  \\
\faLightbulb~Qwen3-32B & \cellcolor[HTML]{CBEDE8}27.44 & \multicolumn{1}{c|}{\cellcolor[HTML]{D7D7D8}2045.45} & \cellcolor[HTML]{9DDCD2}51.93 & \cellcolor[HTML]{CDCDCE}3856.89  \\
\faLightbulb~DeepSeek-R1-0528 & \cellcolor[HTML]{BEE8E1}34.69 & \multicolumn{1}{c|}{\cellcolor[HTML]{BFBFBF}3212.72} & \cellcolor[HTML]{96D9CF}55.41 & \cellcolor[HTML]{BFBFBF}4383.45  \\
\faLightbulb~DeepSeek-v3.2 & \cellcolor[HTML]{C6EBE5}30.31 & \multicolumn{1}{c|}{\cellcolor[HTML]{D9D9DA}1957.45} & \cellcolor[HTML]{8BD5CA}61.11 & \cellcolor[HTML]{C0C0C0}4361.18 \\ \midrule
\multicolumn{1}{c|}{} & \multicolumn{4}{c}{\textbf{\bench-2025}} \\\cline{2-5} 
\multicolumn{1}{c|}{} & \multicolumn{2}{c|}{\textbf{End to End}} & \multicolumn{2}{c}{\textbf{AutoSpec}} \\
\multicolumn{1}{c|}{\multirow{-3}{*}{\textbf{LLMs}}} & {\textbf{pass@1}} & \multicolumn{1}{c|}{\textbf{\#Token}} & {\textbf{pass@1}} & \textbf{\#Token}  \\ \hline
\faRobot~Llama3.3-70B-Instruct & \cellcolor[HTML]{FEFFFF}0.25 & \multicolumn{1}{c|}{\cellcolor[HTML]{FBFBFE}274.00} & \cellcolor[HTML]{A4DED6}17.83 & \cellcolor[HTML]{FCFCFF}1902.79 \\ 
\faRobot~Qwen3-32B & \cellcolor[HTML]{FDFFFF}0.29 & \multicolumn{1}{c|}{\cellcolor[HTML]{FCFCFF}168.90} & \cellcolor[HTML]{AFE3DB}15.56 & \cellcolor[HTML]{FCFCFF}1863.83 \\
\faLightbulb~Qwen3-32B & \cellcolor[HTML]{C9ECE6}7.44 & \multicolumn{1}{c|}{\cellcolor[HTML]{D1D1D2}3836.67} & \cellcolor[HTML]{ABE1D9}16.50 & \cellcolor[HTML]{C8C8C9}5040.30 \\
\faLightbulb~DeepSeek-R1-0528 & \cellcolor[HTML]{8BD5CA}15.73 & \multicolumn{1}{c|}{\cellcolor[HTML]{BFBFBF}5283.57} & \cellcolor[HTML]{B5E5DE}14.44 & \cellcolor[HTML]{BFBFBF}5587.93 \\
\faLightbulb~DeepSeek-v3.2 & \cellcolor[HTML]{DDF3F0}4.65 & \multicolumn{1}{c|}{\cellcolor[HTML]{E2E2E3}2414.12} & \cellcolor[HTML]{8BD5CA}22.56 & \cellcolor[HTML]{CBCBCC}4867.68 \\
\bottomrule
\end{tabular}
}
\label{tab:cost}
\end{table}

Furthermore, the increased token consumption introduced by the reasoning process and AutoSpec primarily stems from output and input tokens, respectively. The former encompasses the model's analysis of the problem, while the latter includes feedback from task decomposition and the examples.

Taking Qwen3-32B as an example for token efficiency comparison. On both subsets, its non-thinking mode within AutoSpec demonstrates lower token consumption than its thinking mode in direct-prompting (2001.10 vs. 2045.45, 1863.83 vs. 3836.67), alongside better pass@1 performance (49.70 vs. 27.44, 15.56 vs. 7.44). This suggests that under constrained token budgets, the benefits derived from the AutoSpec outweigh those from the reasoning. However, results from strong reasoning models, notably DeepSeek-R1, indicate that these two types are currently difficult to compound. 

\refstepcounter{rq}
\vspace{1mm}
\begin{mdframed}[linecolor=gray,roundcorner=12pt,backgroundcolor=gray!15,linewidth=3pt,innerleftmargin=2pt, 
leftmargin=0cm, rightmargin=0cm, topline=false, bottomline=false, rightline=false]
\textbf{RQ\arabic{rq} Summary:} 
All three test-time scaling methods (multiple generation attempts, explicit reasoning chains, and agent workflows) enhance model performance on the ACSL specification generation task. Notably, under a restricted token budget, the agent workflow emerges as a highly cost-effective approach.
\end{mdframed}
\vspace{1mm}

\section{Related Works}\label{sec:related}

Over the past two years, alongside the rapid advancement of LLMs, research focused on leveraging LLMs for specification generation has increasingly emerged. These efforts can be broadly categorized into two areas: the proposal of methodologies for specification writing and the construction of evaluation benchmarks. We mainly focus on automated theorem proving (i.e. ATP) based formal verification (e.g., ACSL, JML, Dafny) because its "auto-active" paradigm offers superior industrial scalability by offloading proof searches to SMT solvers, significantly reducing the human labor required compared to ITPs. This approach aligns perfectly with LLMs' strengths in code-semantic reasoning, enabling an automated, self-correcting feedback loop where models generate formal specifications that can be verified deterministically, providing the most practical path for securing large-scale software systems.

In terms of specification generation, early research focused on evaluating whether LLMs could successfully write loop invariants. Experimental results indicated that by constructing code-loop invariant pairs for fine-tuning, LLMs could be trained to generate correct invariants based on the provided code. Subsequent efforts across various languages integrated the concept of agents, enabling LLMs to collaborate with tools to form structured workflows. Representative works in this area include AutoSpec\cite{wen2024enchantingprogramspecificationsynthesis} for ACSL, SpecGen\cite{ma2025specgenautomatedgenerationformal} for Java Modeling Language~\cite{JML}, and Dafny-synthesis~\cite{Misu_2024,LLM4Dafny} for Dafny\cite{gauci2014dafnystaticallyverifyingfunctional}. These approaches improve the accuracy of model-generated specifications in three ways: first, by utilizing language parsing tools to identify the precise insertion points for specifications within source programs; second, by leveraging verifier feedback for retries, specification deletion, or iterative refinement; third, by using cot instructions and examples to build n-shot prompts. Overall, these tools primarily adopt a workflow-oriented approach, where model outputs and tool calls interact separately with an evaluation sandbox.

In terms of benchmarks, pioneering efforts such as FM-Bench\cite{fmbench} and SpecEval\cite{ma2025specevalevaluatingcodecomprehension} initially established several task formats for specification synthesis. These include translating natural language descriptions into formal specifications or inferring necessary auxiliary specifications based on reference programs and existing assertions. These tasks necessitate a sophisticated understanding of program semantics by LLMs. More recently, a series of benchmarks addressing complex verification scenarios have emerged, such as InvBench\cite{wei2025invbenchllmsaccelerateprogram}, DafnyComp\cite{xu2025localsuccessdoescompose}, and VeriEquivBench\cite{zeng2025veriequivbenchequivalencescoregroundtruthfree}. Furthermore, other studies have shifted their focus toward the joint generation and evaluation of both code and specifications, as exemplified by VerifyThisBench\cite{zeng2025veriequivbenchequivalencescoregroundtruthfree}.

Current benchmarks are progressively introducing more challenging tasks and more precise verification methodologies, thereby placing higher demands on large language models. While this trend is beneficial for the industry's evolution, we contend that a more instructive approach to evaluating specification synthesis is the construction of dynamic datasets characterized by increasing difficulty gradients and chronological variations in publication. The rationale is that different subsets within a dynamic dataset can better characterize the spectrum of a model's capabilities and provide a unified, comparable platform across diverse scenarios, facilitating the prediction and guidance of future model enhancements. Given that the annual SV-COMP competition represents the cutting-edge of formal verification with a gradual difficulty progression, it perfectly satisfies the requirement for real-time assessment. Consequently, we continuously identify and filter suitable C programs from the annual SV-COMP repositories to serve as the source code for our ACSL generation tasks.

\section{Threats to Validity}
There are two major validity threats. In terms of internal validity, a potential threat arises from the models' non-compliance with the form instructions in the prompts, challenging the reliability of the extracted results. For instance, our study explicitly required the specifications to be generated directly or strictly within code blocks, some models still yielded unclosed code blocks, extra explanations, or internal thinking processes. To mitigate this threat, we engineered a more robust specification extraction mechanism. Specifically, we leveraged special "thinking" tokens to filter out the reasoning processes, and stripped away redundant markers and lines containing designated keywords (e.g., "explanation"). 

Regarding external validity, our evaluation is limited to models released before December 21, 2025. We mitigated this by selecting a diverse, representative set of contemporary SOTA models. Additionally, we omitted ultra-large-scale pass@32 experiments for AutoSpec due to prohibitive computational costs. We justify this by noting that AutoSpec exhibits diminishing marginal returns as k increases. Thus, evaluations with smaller k values are sufficient to robustly demonstrate its effectiveness and maintain practical relevance for typical developer environments.

\section{Conclusion}\label{sec:conclusion}

In this paper, we introduce \bench, an evolving benchmark specifically designed for formal specification generation. 

Our empirical evaluation reveals that simply increasing the sampling size could almost double the results. Also, enabling the thinking mode yields substantial improvements, with relative gains ranging from 19.40\% to 2465.52\%. The agentic pipeline further improves performance, but the gain diminishes as the number of sampling attempts increases. Then, we summarize a failure taxonomy, identifying the dominating causes for verification failures. Finally, we conducted a token consumption analysis, confirming the existence of the test-time scaling law and revealing the high cost-effectiveness of the agentic pipeline.

\section*{Acknowledgements}
This work was supported in part by the National Natural Science Foundation of China (Nos. 62372304, 62302375, 62192734), the China Postdoctoral Science Foundation funded project (No. 2023M723736), and the Fundamental Research Funds for the Central Universities.

\newpage

\bibliographystyle{unsrt} 
\bibliography{Tex/reference}

\clearpage

\end{document}